\documentclass[11pt, a4paper, logo, copyright]{googledeepmind}

\usepackage[authoryear, sort&compress, round]{natbib}
\title{\textcolor{golden}{Au}Pair: \textcolor{golden}{Golden} Example Pairs for Code Repair}

\correspondingauthor{mavalankar@deepmind.com}

\author{Aditi Mavalankar}
\author{Hassan Mansoor}
\author{Zita Marinho}
\author{Masha Samsikova}
\author{Tom Schaul}

\affil[1]{Google DeepMind}

\usepackage{times}

\usepackage{amsmath,amsfonts,bm}

\def\eqref#1{equation~\ref{#1}}

\def\1{\bm{1}}

\DeclareMathAlphabet{\mathsfit}{\encodingdefault}{\sfdefault}{m}{sl}
\SetMathAlphabet{\mathsfit}{bold}{\encodingdefault}{\sfdefault}{bx}{n}

\usepackage{graphicx}
\usepackage{hyperref}
\usepackage{wrapfig}
\usepackage{url}
\usepackage{wrapfig}
\usepackage{multirow}
\usepackage{colortbl}
\usepackage{svg}
\usepackage{booktabs}
\usepackage{listings}
\usepackage{textpos}
\usepackage{tcolorbox}
\usepackage{caption}
\usepackage{algorithm}
\usepackage{algpseudocode}
\usepackage{sourcecodepro}
\usepackage[skip=0.5ex]{subcaption}
\usepackage{amsmath}

\definecolor{golden}{rgb}{0.746, 0.5625, 0.0078}
\definecolor{lavender}{rgb}{0.9, 0.9, 0.98}
\definecolor{green0}{rgb}{0.883, 0.955, 0.862}
\definecolor{green1}{rgb}{0.737, 0.896, 0.711}
\definecolor{green2}{rgb}{0.557, 0.816, 0.547}
\definecolor{green3}{rgb}{0.339, 0.712, 0.406}
\definecolor{green4}{rgb}{0.171, 0.582, 0.298}
\definecolor{green5}{rgb}{0.018, 0.443, 0.185}
\definecolor{guessgreen}{rgb}{0.416, 0.659, 0.31}
\definecolor{guessblue}{rgb}{0.24, 0.522, 0.766}

\def\aupair/{\textcolor{golden}{Au}Pair}
\def\aupairs/{\textcolor{golden}{Au}Pairs}

\newcommand*{\inimg}[1]{
    \raisebox{-.2\baselineskip}{
        \includegraphics[
        height=\baselineskip,
        width=\baselineskip,
        keepaspectratio,
        ]{#1}
    }
}

\newcommand*{\inbigimg}[1]{
    \raisebox{-.2\baselineskip}{
        \includegraphics[
        height=18pt,
        width=18pt,
        keepaspectratio,
        ]{#1}
    }
}

\begin{abstract}
Scaling up inference-time compute has proven to be a valuable strategy in improving the performance of Large Language Models (LLMs) without fine-tuning. An important task that can benefit from additional inference-time compute is self-repair; given an initial flawed response, or guess, the LLM corrects its own mistake and produces an improved response, or fix. We leverage the in-context learning ability of LLMs to perform self-repair in the coding domain. The key contribution of our paper is an approach that synthesises and selects an ordered set of \textcolor{golden}{golden} example \emph{pairs}, or \aupairs/, of these initial guesses and subsequent fixes for the corresponding problems. Each such \aupair/ is provided as a single in-context example at inference time to generate a repaired solution. For an inference-time compute budget of $N$ LLM calls per problem, $N$ \aupairs/ are used to generate $N$ repaired solutions, out of which the highest-scoring solution is selected as the final answer. The underlying intuition is that if the LLM is given a different example of fixing an incorrect guess each time, it can subsequently generate a diverse set of repaired solutions. Our algorithm selects these \aupairs/ in a manner that maximises complementarity and usefulness. We demonstrate the results of our algorithm on 5 LLMs across 7 competitive programming datasets for the code repair task. Our algorithm yields a significant boost in performance compared to best-of-$N$ and self-repair, and also exhibits strong generalisation across datasets and models. Moreover, our approach shows significantly stronger scaling with inference-time compute budget compared to baselines.
\end{abstract}

\begin{document}

\maketitle

\section{Introduction}

Recent progress in the field of Large Language Models (LLMs) has resulted in models that keep getting better at generating responses to user queries. When providing these already powerful models with more inference-time compute---increasing number of LLM calls---methods that sample different responses and then select the best among them, such as best-of-$N$~\citep{best-of-n} or self-consistency~\citep{wang2023selfconsistencyimproveschainthought}, have shown clear benefits. While these approaches are more breadth-focused, another way to leverage inference time compute is to improve or \emph{repair} the LLM's initial \emph{guesses} by generating better \emph{fixes}~\citep{olausson2024self}. We propose combining the benefits of both these approaches to generate a wide set of repaired solutions for poor initial LLM responses, out of which the highest-scoring solution is the final answer.

\begin{figure}[t]
    \includegraphics[width=\linewidth,trim={0.5cm, 0cm, 5.5cm, 0cm}, clip]{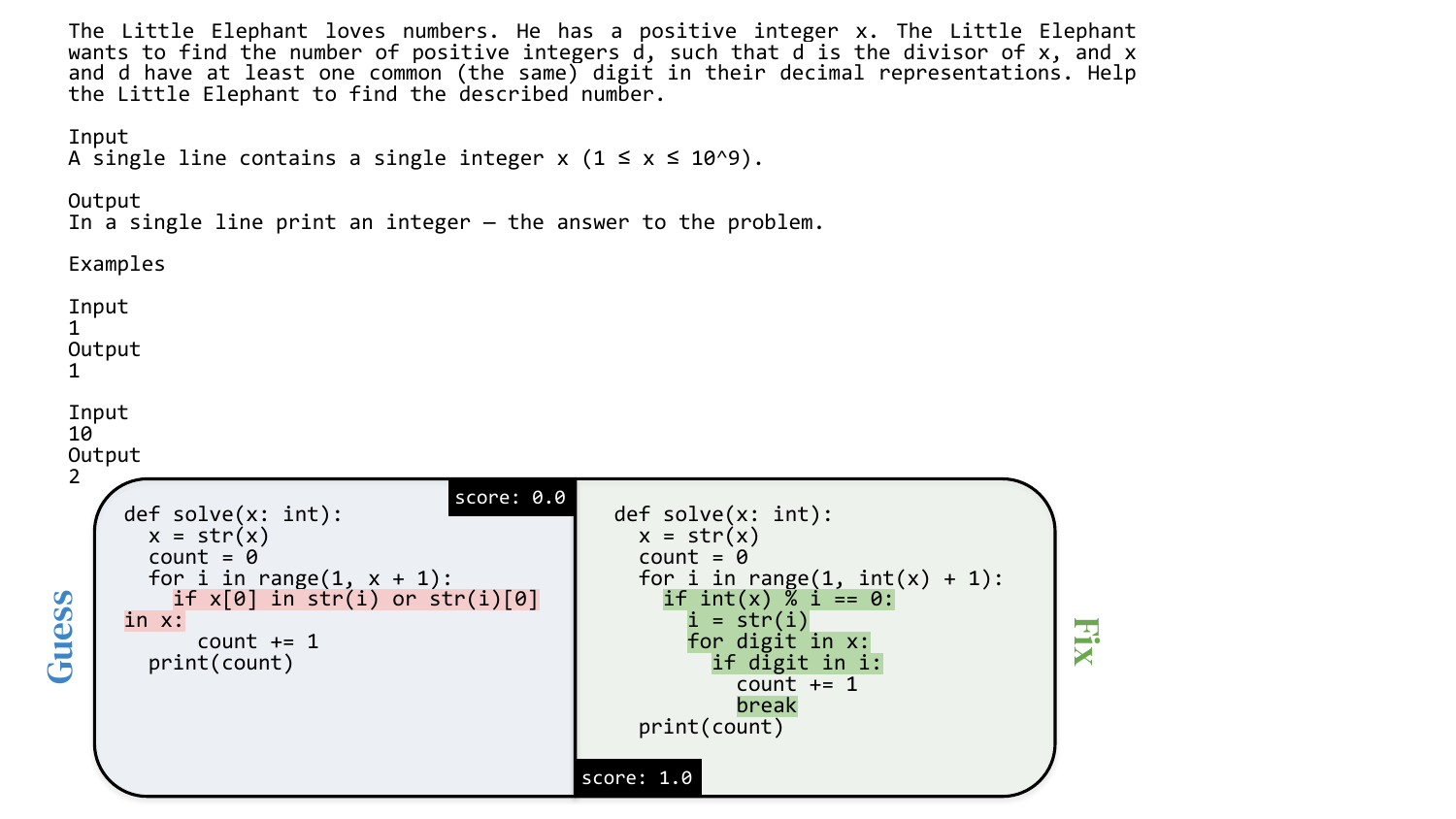}
    \caption{\textbf{An example \aupair/} \inbigimg{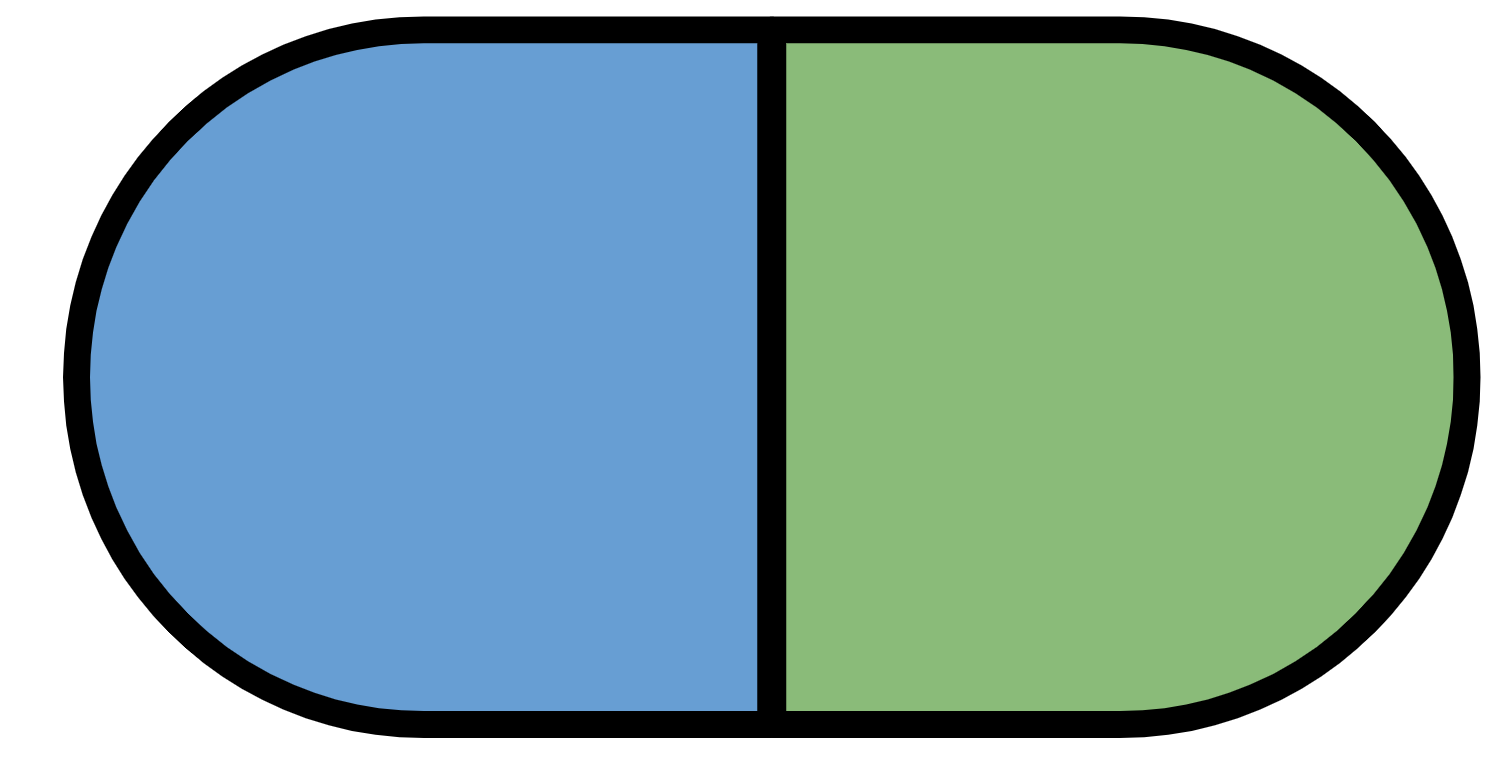}: The LLM-generated guess and fix, along with their respective scores for the corresponding CodeForces problem (problem description at the top). The guess checks only the first digit for every single number leading up to the input. The fix corrects the logic by iterating over the \emph{divisors} of the input, and checking for an intersection over \emph{all} digits with the input. To provide this \aupair/ in context at inference time, the problem description, guess, and fix, are concatenated as described in \S\ref{fig:repair_prompt}.}
    \label{fig:example_aupair}
\end{figure}

To generate a wide range of repaired solutions for each initial LLM response, we exploit the in-context learning capability exhibited by LLMs. The main contribution of our paper is an algorithm that, given an inference-time compute budget of $N$ LLM calls, produces an ordered set of up to $N$ \textcolor{golden}{golden} example pairs, or \aupairs/\footnote{The name \aupair/ is a coupling of \textcolor{golden}{Au}, the chemical symbol for gold, and Pair, jointly referring to \textcolor{golden}{golden} pairs that are produced by our algorithm. The high-level interpretation is that like an "au pair", the approach guides the LLM towards better behaviour.}. Each such \aupair/ consists of the initial guess and the consequent fix for the corresponding coding problem, along with their respective unit test scores. An example \aupair/ is illustrated in Fig.~\ref{fig:example_aupair}. At inference time, the contents of an \aupair/ are concatenated as described in \S~\ref{sec:prompting}, and provided as a 1-shot example to generate an improved solution or fix for the test problem. This is done for each of the $N$ \aupairs/; of all the fixes generated, the one that gets the highest score on the unit tests is selected.

A core ingredient of our proposed algorithm involves the selection of these \aupairs/. We propose a submodular approach that selects \aupairs/ based on the ability of each pair to solve different problems in a held-out validation set. Since the \aupairs/ are selected such that each subsequent \aupair/ solves a different set of problems than the ones solved by its predecessor \aupairs/, by design, we get \emph{complementary} \aupairs/. Also, as the list of \aupairs/ is constructed by taking the greedy pair at each step, only those pairs that lead to an increase in the fix scores are selected, resulting in \emph{useful} \aupairs/.

In this paper, we address the code repair task: given a coding problem, an initial guess which is LLM-generated code, and a set of test cases that are used only to evaluate the correctness of the generated code, the LLM has to generate an improved fix for the problem. We show that the fixes generated by \aupair/ are significantly more \emph{useful} and \emph{diverse} than those generated using best-of-$N$ (\S\ref{sec:results}) for the same inference-time compute budget. We also show that \aupair/ outperforms self-repair~\citep{olausson2024self}, which generates intermediate verbal feedback before generating repaired solutions for a problem. \\

The key contributions of our paper are the following:
\begin{itemize}
    \item An inference-time {\bf algorithm}, which constructs a \textcolor{golden}{golden} set of code repair examples, or \aupairs/, that boost performance significantly when used as in-context examples (\S\ref{sec:approach}).
    \item {\bf Reliably outperforming} best-of-$N$~\citep{best-of-n} and self-repair~\citep{olausson2024self} across 5 different models: Gemini-1.5-Pro, GPT-4o-mini, Gemini-1.5-Flash, Gemma-27B, and Gemma-9B (\S\ref{sec:in_dist_performance}).
    \item Strong {\bf scaling} performance with inference time compute, with far less diminishing returns than best-of-$N$ and self-repair (\S\ref{sec:inference_compute_scaling}). 
    \item Robust out-of-distribution {\bf generalisation}, across both model sizes and datasets (\S\ref{sec:OOD}).
\end{itemize}

\section{Approach}
\label{sec:approach}

\begin{figure}[t]
    \centering
    \includegraphics[trim={1cm 3cm 5cm 2.5cm}, scale=0.8, clip, width=0.95\linewidth]{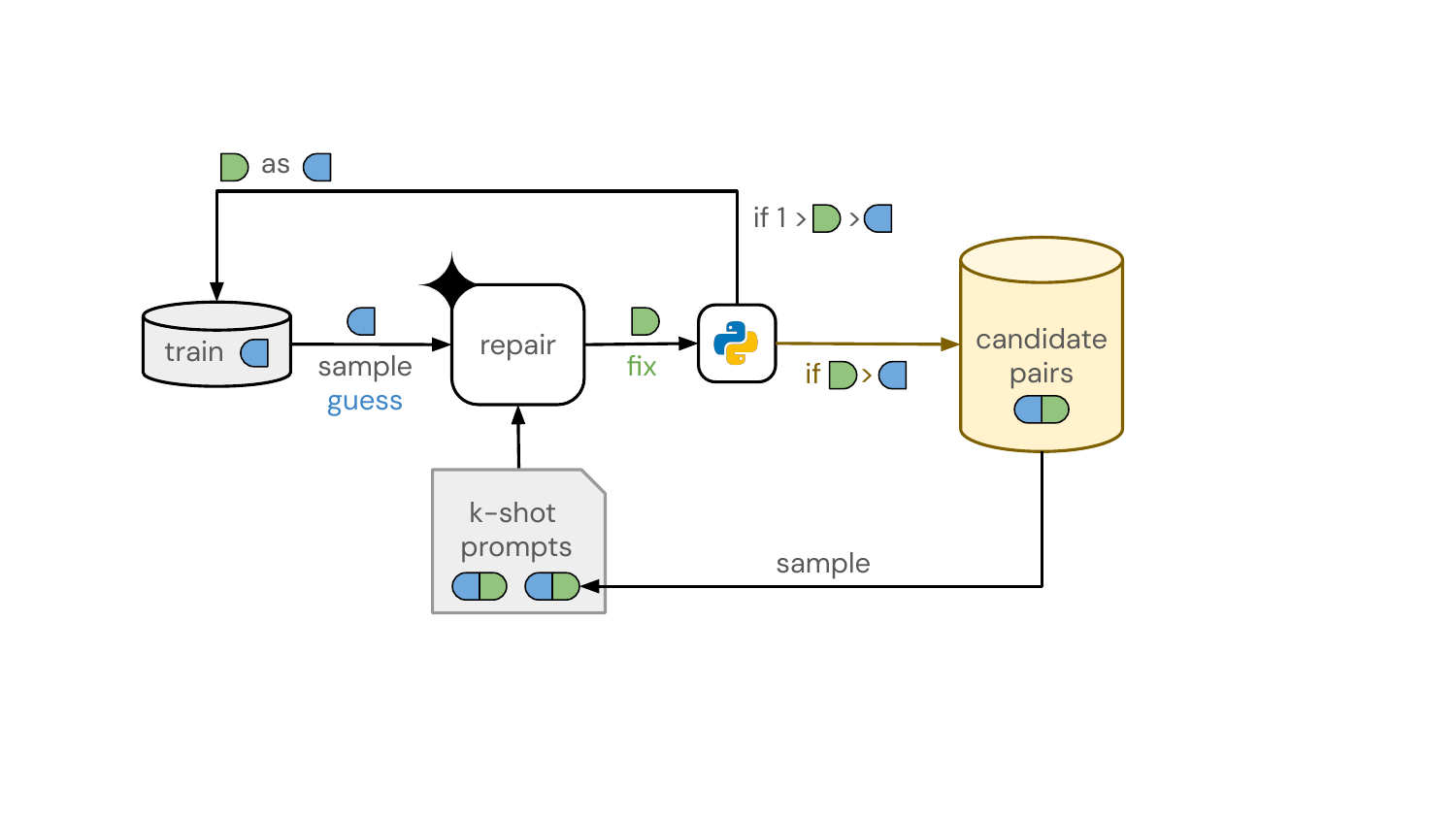}
    \caption{\textbf{Pair Generation:} This phase includes collecting a large set $\mathcal{C}$ of guesses for coding problems \inimg{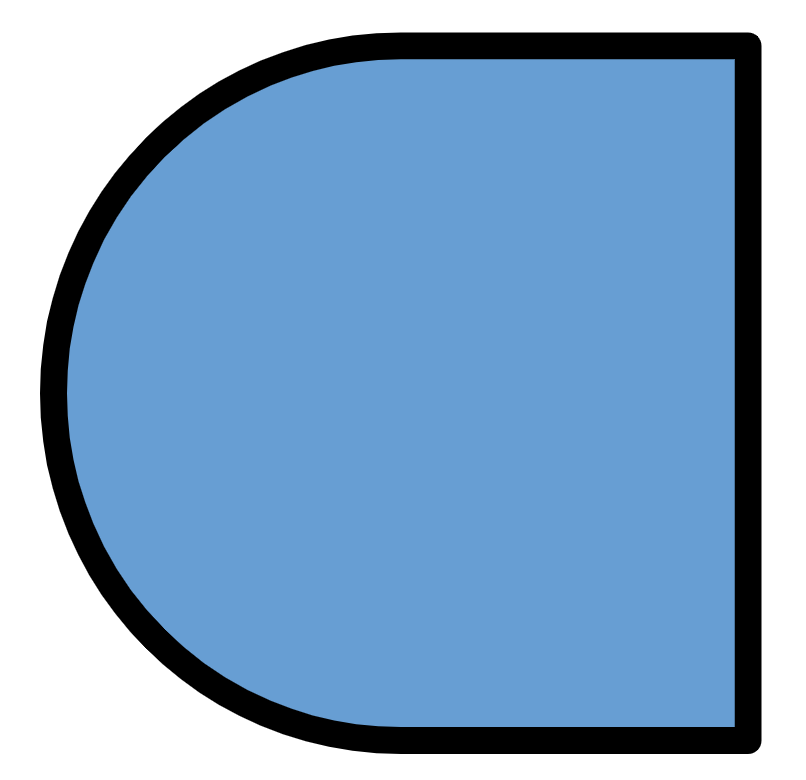} and their fixes \inimg{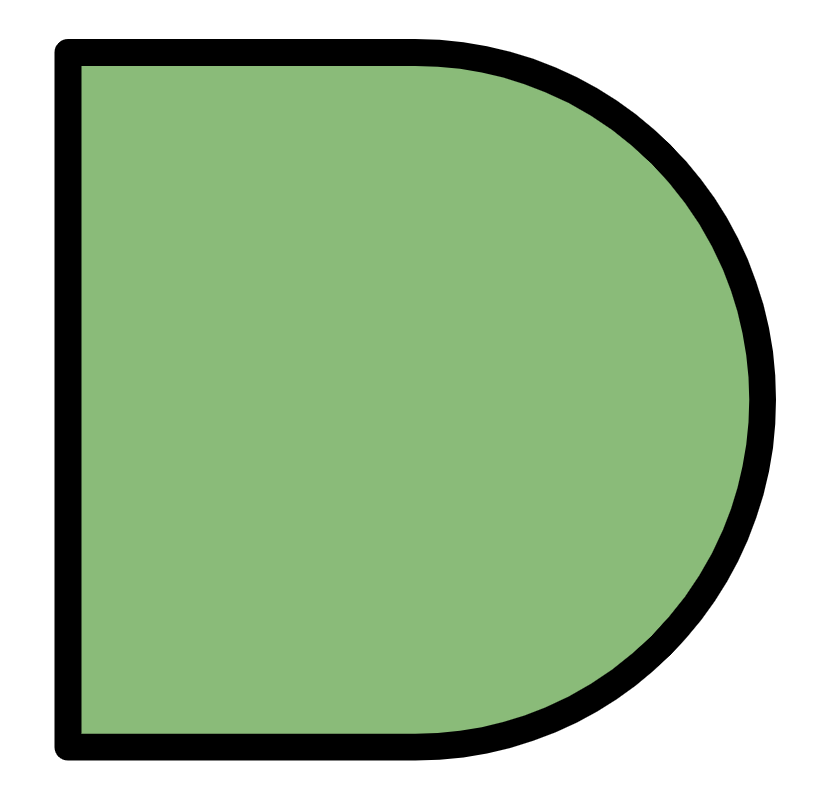}, yielding candidate pairs \inbigimg{figures/pair.png} that will later be used to get \aupairs/. At each step, a problem with its guess is sampled from the training dataset, and used in conjunction with $k$ randomly sampled pairs from the candidate pair buffer to compose a $k$-shot prompt. This prompt is then passed through an LLM to generate a fix, which is evaluated on the unit tests by running the Python interpreter and computing its test pass rate. If this fix is better than the guess, this (guess, fix) pair is added to the set of candidate pairs. Any improved but imperfect fix is also added as a new guess to the training dataset. See \S\ref{sec:phase1} for more details.}
    \label{fig:phase1}
\end{figure}

The goal of our proposed approach in the coding domain is to improve code repair performance on unit tests at inference time, by curating a list of pairs that can be provided as in context examples. The code repair prompt includes an in-context example, followed by a text description of the problem to solve and the initial \emph{guess} generated by the LLM. The LLM then generates a revision, or a \emph{fix} that improves performance on the unit tests for that problem. In the prompt, we also include the scores achieved by the guess and fix on the unit tests, but no additional execution feedback.\footnote{The repair prompt is composed using the prompting strategy shown in Fig.~\ref{fig:repair_prompt}.}

In order to disentangle repair performance from the quality of initial guesses, we first curate composite datasets consisting of initial guesses for all the coding problems. Given a dataset consisting of problems and their corresponding tests, we generate an initial \emph{guess} for each problem and compute its score on the unit tests. If the guess passes all the unit tests for that problem correctly, no further improvement is required and we discard that problem. If not, we add this guess along with its corresponding score and problem as a datapoint to our curated dataset. This dataset is then divided into training, validation, and test datasets. We use the training dataset $\mathcal{D}_\text{train}\equiv \mathcal{D}$ for pair generation (Fig.~\ref{fig:phase1}), and the validation dataset $\mathcal{D}_\text{val}$ for \aupair/ extraction. The test dataset is used in the final testing phase only $\mathcal{D}_\text{test}$.

Following the creation of the training, validation, and test datasets, we now discuss our approach, which consists of two main phases: 1) Pair Generation (\S\ref{sec:phase1}), and 2) \aupair/ Extraction (\S\ref{sec:phase2}).

\subsection{Phase 1: Pair Generation}
\label{sec:phase1}

In this phase, we generate a large set $\mathcal{C}$ of pairs that are potential candidates for our final set of \aupairs/. This is done in the following manner: a problem along with its initial guess is sampled from the training dataset $\mathcal{D}$. The LLM generates a fix for this guess. If this generated fix has a higher score on the unit tests for that problem than the initial guess, this guess-fix pair is added to $\mathcal{C}$. Furthermore, if this fix is imperfect, i.e. it does not pass all the unit tests, it becomes a potential guess with further scope for improvement, so it is added as a new guess to the training dataset $\mathcal{D}$. This is repeated several times to collect a large set of such candidate pairs.\footnote{Since the aim is to collect a large set of pairs, we want the LLM to generate a wide variety of fixes. For this, we compose a $k$-shot prompt for repair, in which the $k$ in-context example pairs are randomly sampled from the existing set $\mathcal{C}$ for each sampled problem. As the LLM generates more fixes, $\mathcal{C}$ gets populated, subsequently resulting in more diverse prompts.} A visual illustration of this phase is provided in Fig.~\ref{fig:phase1}.

While we include the pair generation phase for completeness, it is important to note that in several other domains, paired data may already be available. In such cases, the set of candidate pairs $\mathcal{C}$ that we curate in this step, can simply be replaced by the given paired data.

\subsection{Phase 2: \aupair/ Extraction}
\label{sec:phase2}

\begin{figure}[t]
    \centering
    \includegraphics[trim={1cm 2cm 2cm 2.5cm}, clip, width=0.95\linewidth]{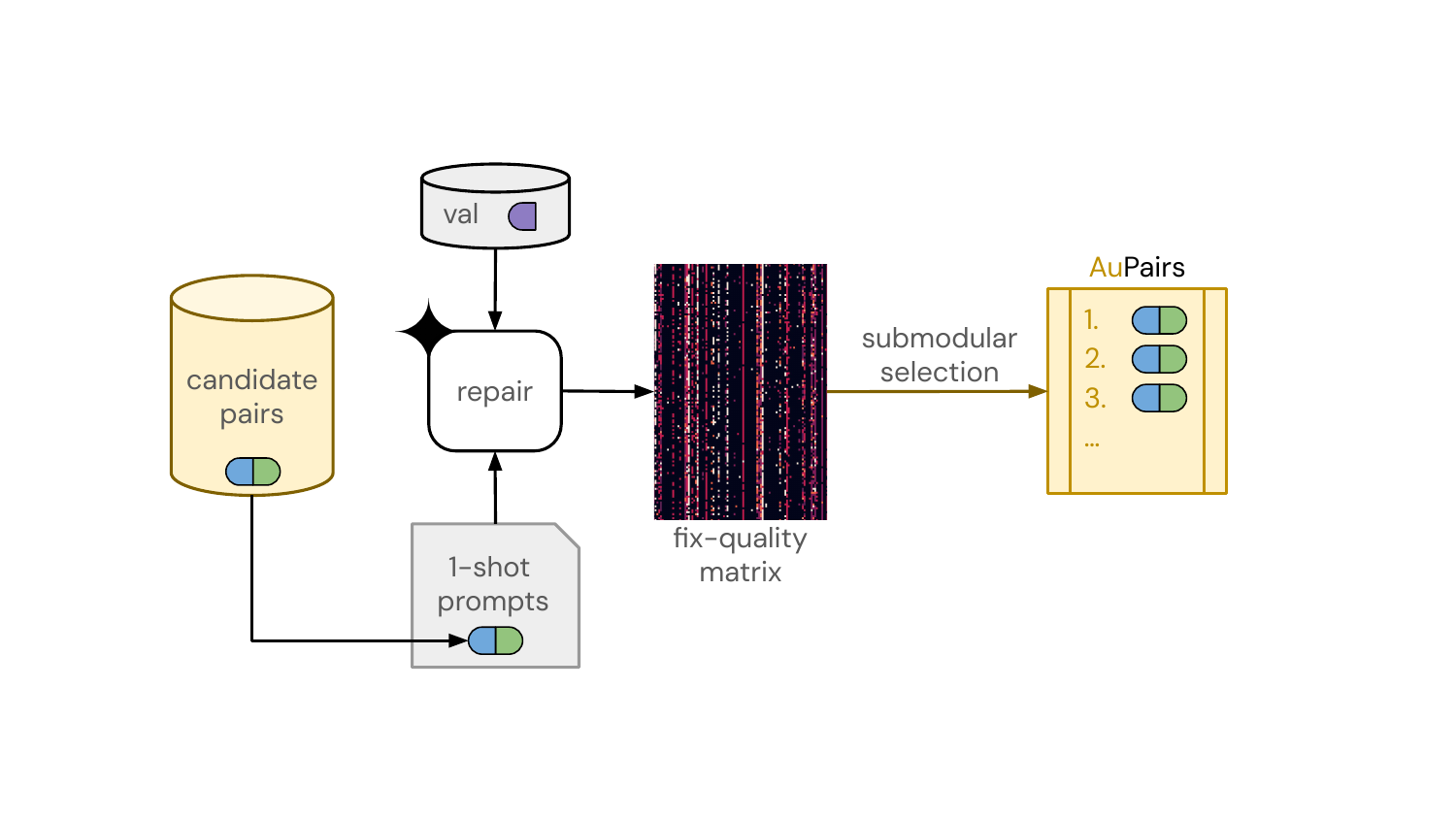}
    \caption{\textbf{\aupair/ Extraction:} given a large set of candidate pairs \inbigimg{figures/pair.png}, the next step is to extract \aupairs/ from them. For this, each pair \inbigimg{figures/pair.png} is provided as a 1-shot in-context example in the prompt for each problem and its guess \inimg{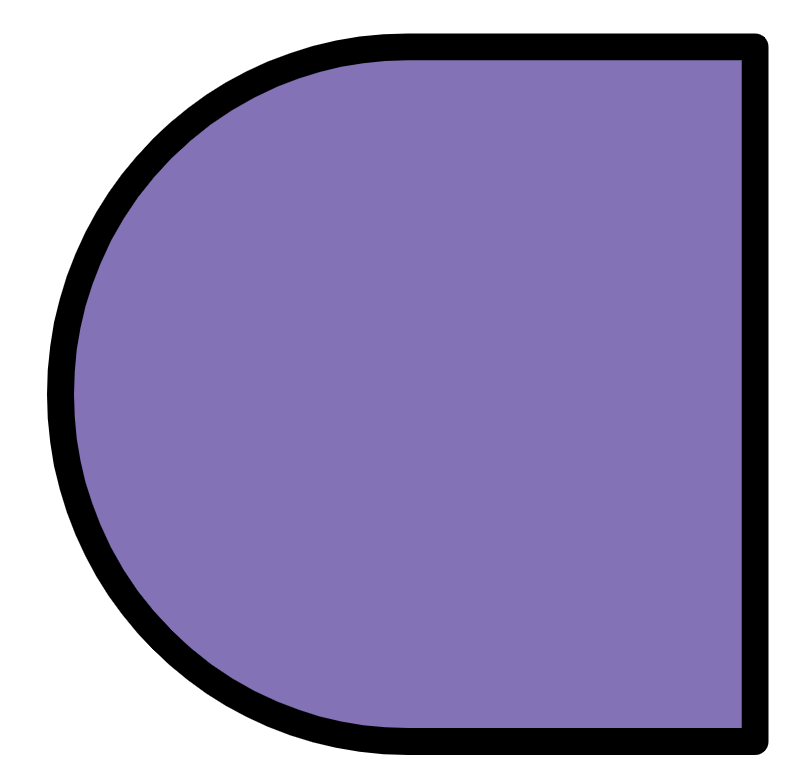} from the validation dataset. These prompts are then passed to the LLM which generates fixes that are evaluated on the corresponding unit tests to populate a fix-quality matrix, as described in Algorithm~\ref{alg:fix-quality}. Following this, a submodular selection mechanism is applied on this fix-quality matrix to obtain the list of \aupairs/, as described in Algorithm~\ref{alg:submodular}.}
    \label{fig:phase2}
\end{figure}

Now that we have a large set $\mathcal{C}$ of candidate pairs, the next step is to determine which of these will actually help boost performance, i.e., which of these are \aupairs/. We do this in a \emph{submodular} fashion by making use of the validation dataset $\mathcal{D}_\text{val}$. For each pair-problem combination $(\bm{c}_i, \bm{x}_j) \in \mathcal{C} \times \mathcal{D}_\text{val}$, we build a 1-shot prompt $\bm{p}$ using the prompting strategy described in \ref{fig:repair_prompt}. This 1-shot prompt $\bm{p}$ is given as input to the LLM, which generates a fix for the given problem $\bm{x}_j$. The fix generated by the LLM is then evaluated on the unit tests and stored in the fix quality matrix $\bm{M} \in \mathbb{R}^{|\mathcal{C}|\times|\mathcal{D}_\text{val}|}$ at index $(i, j)$. This part of \aupair/ extraction is outlined in Algorithm~\ref{alg:fix-quality}.

\begin{tabular}{cc}
\begin{minipage}[c]{0.47\linewidth}     
\begin{algorithm}[H]
\begin{algorithmic}[1]
\Require
$\left\{\begin{array}{ll}
\text{LLM} & \text{large language model} \\
\mathcal{C} & \text{candidate pairs} \\
\mathcal{D_\text{val}} & \text{validation dataset}\\
\text{score} & \text{code eval function} \\
\end{array}
\right.$
\State init fix quality matrix $\bm{M} \gets \mathbf{0}^{|\mathcal{C}| \times |\mathcal{D_\text{val}}|}$
\For{pair $\bm{c}_i$, problem $\bm{x}_j \in \mathcal{C} \times \mathcal{D}_\text{val}$}
\State build 1-shot prompt: $\bm{p} \gets \bm{c}_i \mathbin\Vert \bm{x}_j$
\State generate fix: $\bm{\hat{y}} \gets \text{LLM}(\bm{p})$
\State evaluate fix: $\bm{M}_{i,j} \gets \text{score}(\bm{\hat{y}})$
\EndFor
\Statex \Return{$\bm{M}$}
\end{algorithmic}
\caption{Fix quality matrix computation}
\label{alg:fix-quality}
\end{algorithm}
\end{minipage}
&
\begin{minipage}[c]{0.47\linewidth}
\begin{algorithm}[H]
\begin{algorithmic}[1]
\Require
$\left\{\begin{array}{ll}
\bm{M} & \text{fix quality matrix} \\
\mathcal{C} & \text{candidate pairs} \\
\epsilon & \text{tolerance} \\
\end{array}
\right.$
\State initialise \aupairs/ $\mathcal{A} \gets []$
\Repeat
\State per-pair scores: $\bm{\bar{m}} \gets \text{row-mean}(\bm{M})$
\State get best pair: $\bm{c_k} \gets \text{argmax}_{\mathcal{C}} \bm{\bar{m}}$
\State append to \aupairs/: $\mathcal{A} \gets \mathcal{A} + \bm{c_k}$
\State update $\bm{M} \gets \text{clip}(\bm{M} - \bm{M}_k, 0, 1)$
\Until $\text{max}({\bm{\bar{m}}}) < \epsilon$
\Statex \Return{$\mathcal{A}$}
\end{algorithmic}
\caption{Submodular \aupair/ extraction}
\label{alg:submodular}
\end{algorithm}
\end{minipage}
\end{tabular}

Next, we use this fix quality matrix $\bm{M}$ to extract the \aupairs/ by taking the following steps: 
1) Select the pair that gets the highest mean score across all problems in $\mathcal{D}_\text{val}$, say $\bm{c}_k$, and add it to the list of \aupairs/ $\mathcal{A}: \mathcal{A} \gets \mathcal{A} \cup \bm{c}_k$. This is a greedy way of selecting the best pair given all previous \aupairs/ and produces an ordered set of \aupairs/. 2) Subtract the row score $\bm{M}_k$ (i.e. score on all the problems in $\mathcal{D}_\text{val}$) of this newly added pair from all the rows in the fix quality matrix with an update: $\bm{M} - \bm{M}_k$. This ensures that redundant \aupairs/ are not produced by the approach. The updated fix quality matrix is clipped to $(0, 1)$ since any negative value in the matrix $\bm{M}$, say $M_{i,j}$, implies that the problem $\bm{x}_j$ cannot be improved further by pair $\bm{c}_i$. Without clipping, we would not get an accurate estimate of the improvement in the next step of submodular extraction. 3) Repeat this process until improvement falls beyond a certain tolerance $\epsilon$.

This process of iteratively constructing the set of \aupairs/ ensures that they improve performance on disjoint parts of the problem space. The \aupairs/ that we obtain from this phase are used in the same manner at inference time, as 1-shot examples, to improve code repair performance. The compute budget $N$ determines the number of \aupairs/ that we can use at inference time. Since the \aupairs/ form an ordered set, the first $N$ \aupairs/ are used for a compute budget of $N$ LLM calls. The final solution for each problem is the one that passes the most test cases, among all generated solutions. This submodular extraction of \aupairs/ is shown in Algorithm~\ref{alg:submodular}. Fig.~\ref{fig:phase2} has a joint diagram depicting fix quality matrix computation and submodular \aupair/ extraction.

\section{Experiments}
\label{sec:results}

\textbf{Datasets:} We use 7 datasets that contain problems and test cases from competitive programming contests: 1) CodeForces (8.8k problems), 2) AtCoder (1.3k problems), 3) HackerEarth (1.2k problems), 4) CodeChef (768 problems), 5) LiveCodeBench (400 problems),  6) CodeJam (180 problems), and 7) Aizu (2.2k problems)~\citep{codecontests, jain2024livecodebench}. We choose CodeForces and AtCoder, separately, for in-distribution testing, and use the rest exclusively for out-of-distribution testing. Our training / validation / test split proportions for the CodeForces and AtCoder datasets are $37.5 / 12.5 / 50\%$. Some datasets have difficulty levels as part of the problem; for those datasets we maintain the same stratified distribution of questions in the training, validation, and test datasets.

\textbf{Models:} We demonstrate the superior code repair capability of \aupair/ on 5 different models: Gemini-1.5-Pro, GPT-4o-mini, Gemini-1.5-Flash, Gemma-27B, and Gemma-9B. In addition to using these models for dataset curation and pair generation, we also look at the transfer capabilities of our method with respect to different models in \S\ref{sec:model_transfer_analysis}.

\textbf{Evaluation:} We perform two types of evaluation: in-distribution and out-of-distribution. For in-distribution evaluation, we use the test split from the same dataset as the one used for pair generation and \aupair/ extraction. This ensures that the format of questions and test cases in the test questions matches that of the \aupairs/. Out-of-distribution evaluation uses a different coding dataset; this means that the test samples have different format of questions, difficulty, types of problems and test cases than the \aupairs/. Another axis of out-of-distribution evaluation that we study is the model axis: we report the performance obtained using \aupairs/ produced by a different model than the one used at inference time.

\textbf{Metrics:} Our primary metric is the commonly used test pass rate, also called test case average~\citep{hendrycks2021measuring,10.1145/3697010,wu2024benchmarking}, which we compute as the average percentage of test cases passed. In our setting, since we choose the best out of $N$ responses generated by the LLM, the test pass rate for a test dataset with $P$ problems is calculated as: 

$$\text{test pass rate} = \frac{1}{P}\sum_{p=1}^{P}\max_{i \in \{1, \ldots, N\}} \frac{1}{|T_p|} \sum_{j=1}^{|T_p|} \mathbb{1} \{ \texttt{eval}(\text{code}_{p, i}, T_{p, j}) == \texttt{pass} \}$$

where $T_p$ refers to the unit tests for problem $p$, and $\text{code}_{p, i}$ is the code generated by the LLM for problem $p$ in the $i^{\text{th}}$ LLM call. The innermost loop computes the percentage of unit tests passed by the LLM output $\text{code}_{p, i}$. Following this, we select the code output that has the highest value for percentage of unit tests passed, which is evident from the $\max$ operation over all LLM calls $i \in \{ 1, \ldots, N \}$. The outermost loop averages this across all the problems in the test dataset.

We also report the results of strict accuracy~\citep{hendrycks2021measuring}, which is the percentage of generated solutions that pass all test cases (definition and results in Appendix \S\ref{sec:strict_accuracy}). Note that we cannot provide fair results for the pass@$k$ metric because the pass@$k$ metric makes an assumption that all $k$ responses from the LLM are i.i.d. generated, whereas our approach produces fixes in a specific order. In our case, the first \aupair/ is more useful than the second, which is more useful than the third, etc; so their success probabilities monotonically decrease as a function of $k$. 

\textbf{Baselines:} We compare the effectiveness of our proposed approach with best-of-$N$~\citep{best-of-n} and self-repair~\citep{olausson2024self}. Best-of-$N$ is a strong baseline to improve model performance by allowing multiple LLM calls at inference time. Of the $N$ generated responses, the highest-scoring response is selected. To ensure the sampling of high-quality diverse responses in best-of-$N$, we set the temperature to 1.0~\citep{temperature}. Self-repair, on the other hand, uses the compute budget of $N$ LLM calls to either generate verbal feedback or repaired code. Our compute budget is $N = 32$, of which 4 LLM calls are used to generate verbal feedback and 7 LLM calls to generate repaired code for each verbal feedback.

The remainder of this section will discuss a plethora of empirical results, on overall and ablated performance (\S\ref{sec:in_dist_performance} and~\ref{sec:pairs_vs_aupairs}), scalability and generalisation (\S\ref{sec:inference_compute_scaling} to~\ref{sec:model_transfer_analysis}), and diversity (\S\ref{sec:aupair_analysis_diversity} to~\ref{sec:categories}).

\begin{figure}[t]
    \centering
    \includegraphics[width=0.45\linewidth]{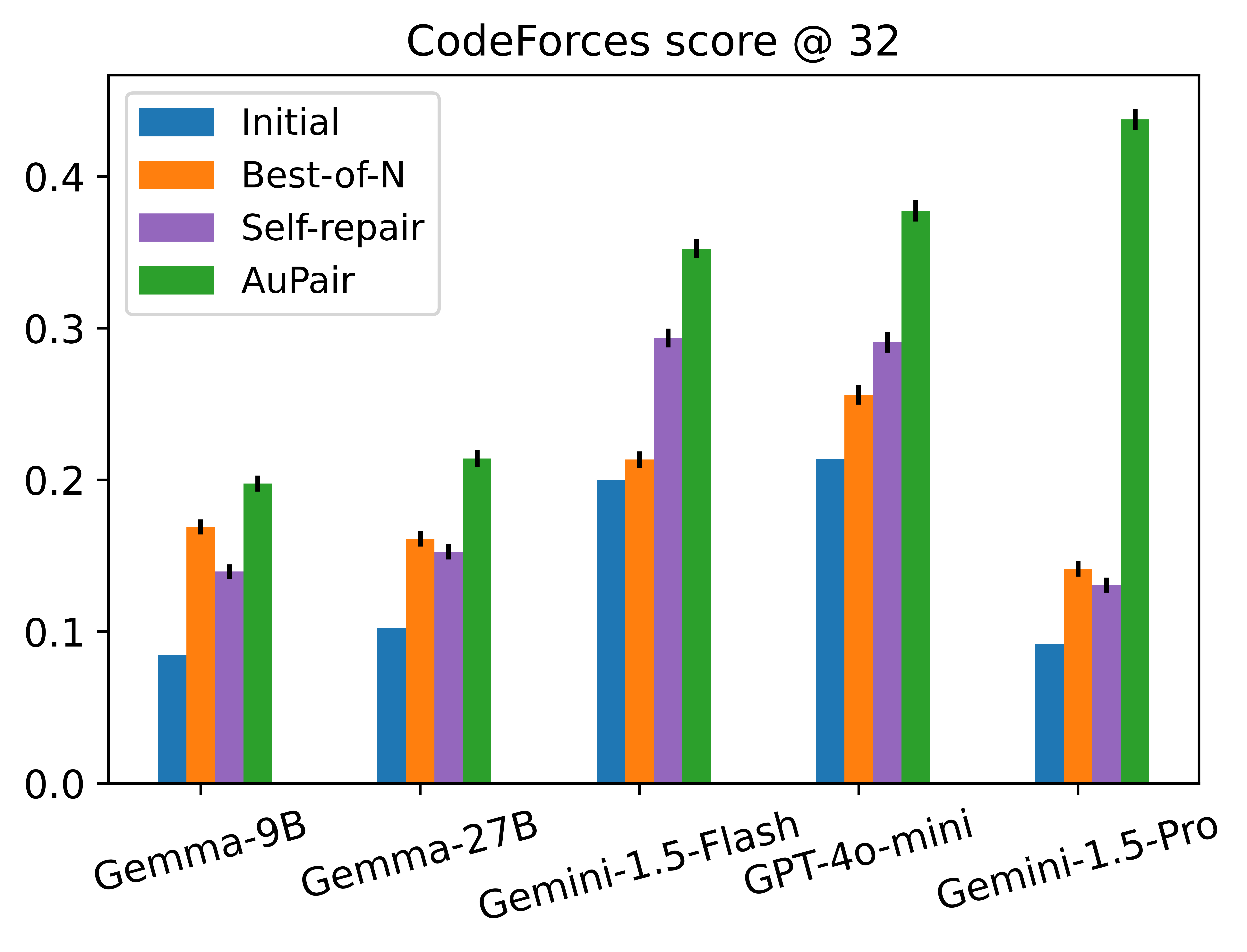}
    \includegraphics[width=0.45\linewidth]{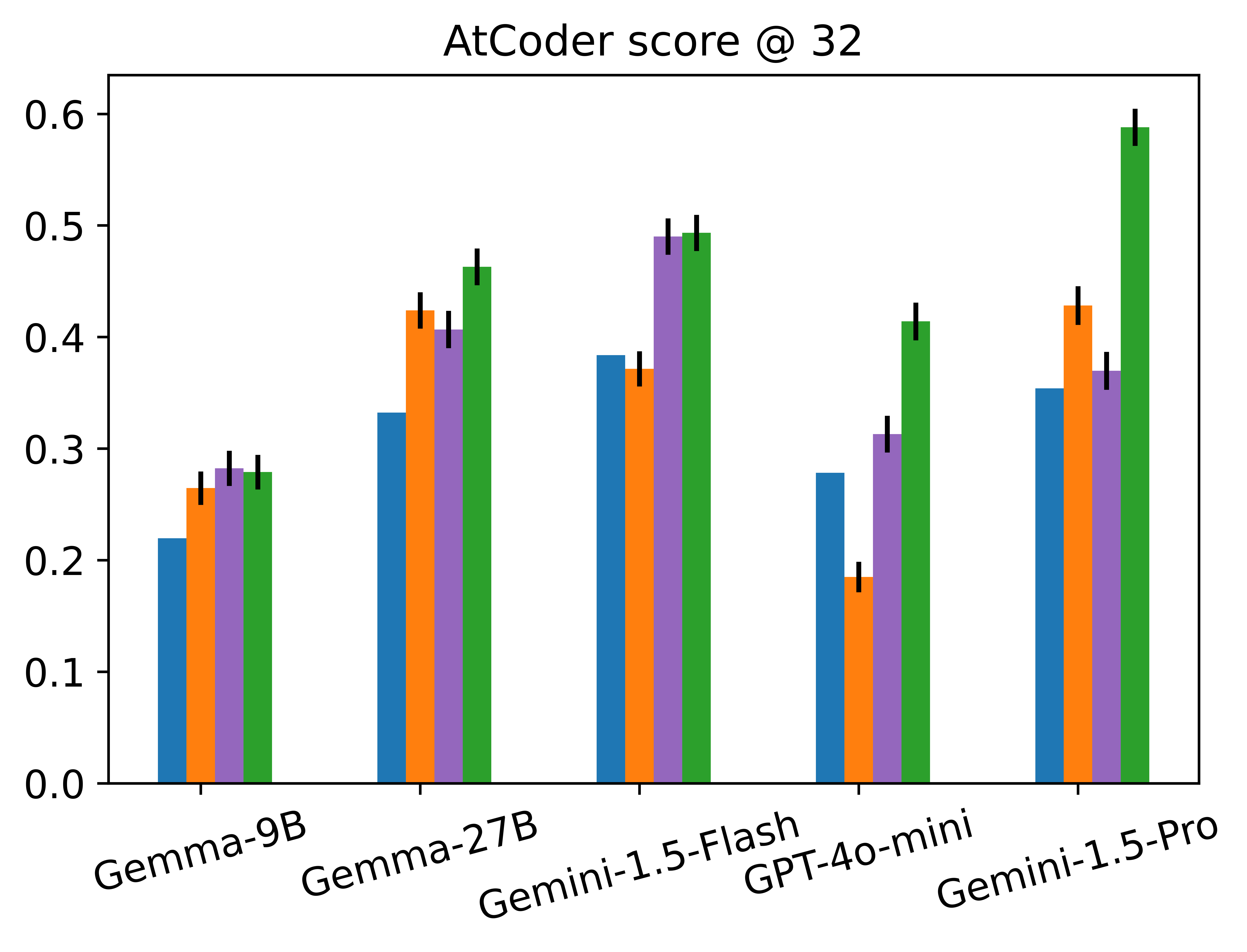}
    \caption{\textbf{In-distribution code repair performance:} with $N = 32$ LLM calls at inference time and the same train / val / test data distribution, we compute the test pass rate. The same model is used for generating the initial guesses and fixes and the \aupair/ extraction. CodeForces (left, 8.8k problems) and AtCoder (right, 1.3k problems), see~\S\ref{sec:in_dist_performance} for more details.}
    \label{fig:in_dist_performance}
\end{figure}

\subsection{Significantly Boosted Code Repair Performance}
\label{sec:in_dist_performance}

The first step to assess code repair performance is to measure
\emph{in-distribution} performance; namely generating and selecting \aupairs/ on the training and validation sets that match the test dataset, and using the same model for evaluation as \aupair/ construction. We do this for 2 datasets (CodeForces and AtCoder) and all 5 models. Fig.~\ref{fig:in_dist_performance} shows the resulting comparison between the best-of-$N$ and self-repair baselines and \aupair/, for a budget of $N = 32$ LLM calls at inference time.\footnote{Since our algorithm yields a variable number of \aupairs/, for smaller datasets with fewer generated pairs, the total number of \aupairs/ can be less than 32. To have a fair comparison in that case, we set the same compute budget $N$ for best-of-$N$ and self-repair. This is the case for AtCoder (Fig.~\ref{fig:in_dist_performance}, right), where our algorithm yields 14, 15, and 27 \aupairs/ for Gemma-9B, GPT-4o-mini, and Gemma-27B respectively. So the corresponding baseline results also use a matching compute budget of 14, 15, and 27 LLM calls respectively.} \aupair/ is clearly superior to best-of-$N$ and self-repair (matching in a few cases) on all models and datasets, sometimes by wide margins. This clearly establishes that our proposal of providing a different in-context example of code repair in each LLM call can significantly boost performance.

An interesting side-result is visible in initial performance, i.e., the performance of the initial responses of the LLMs to the problems, which have to then be repaired. Gemini-1.5-Pro, despite being a superior model to Gemini-1.5-Flash, shows worse initial performance. Since the code generated has certain conditions that allow successful execution, we observe that many initial guesses of generated code fail because they do not obey these conditions (see Appendix~\S\ref{sec:code_execution}). In such cases, code repair with either best-of-$N$ or self-repair is unlikely to give us high boost in performance since the initial solution is badly formatted. This is one clear case where having an \aupair/ in context significantly improves performance. As a result, using \aupairs/ in conjunction with high performing models leads to large performance improvements despite poor initial performance, as we can see for both CodeForces and AtCoder with the Gemini-1.5-Pro model in Fig.~\ref{fig:in_dist_performance}. This also mitigates the need for more sophisticated prompt engineering to a large extent.

\subsection{Selection Matters: \aupairs/ are More Effective than Random Pairs}
\label{sec:pairs_vs_aupairs}

\begin{figure*}
    \centering
    \includegraphics[width=0.45\linewidth]{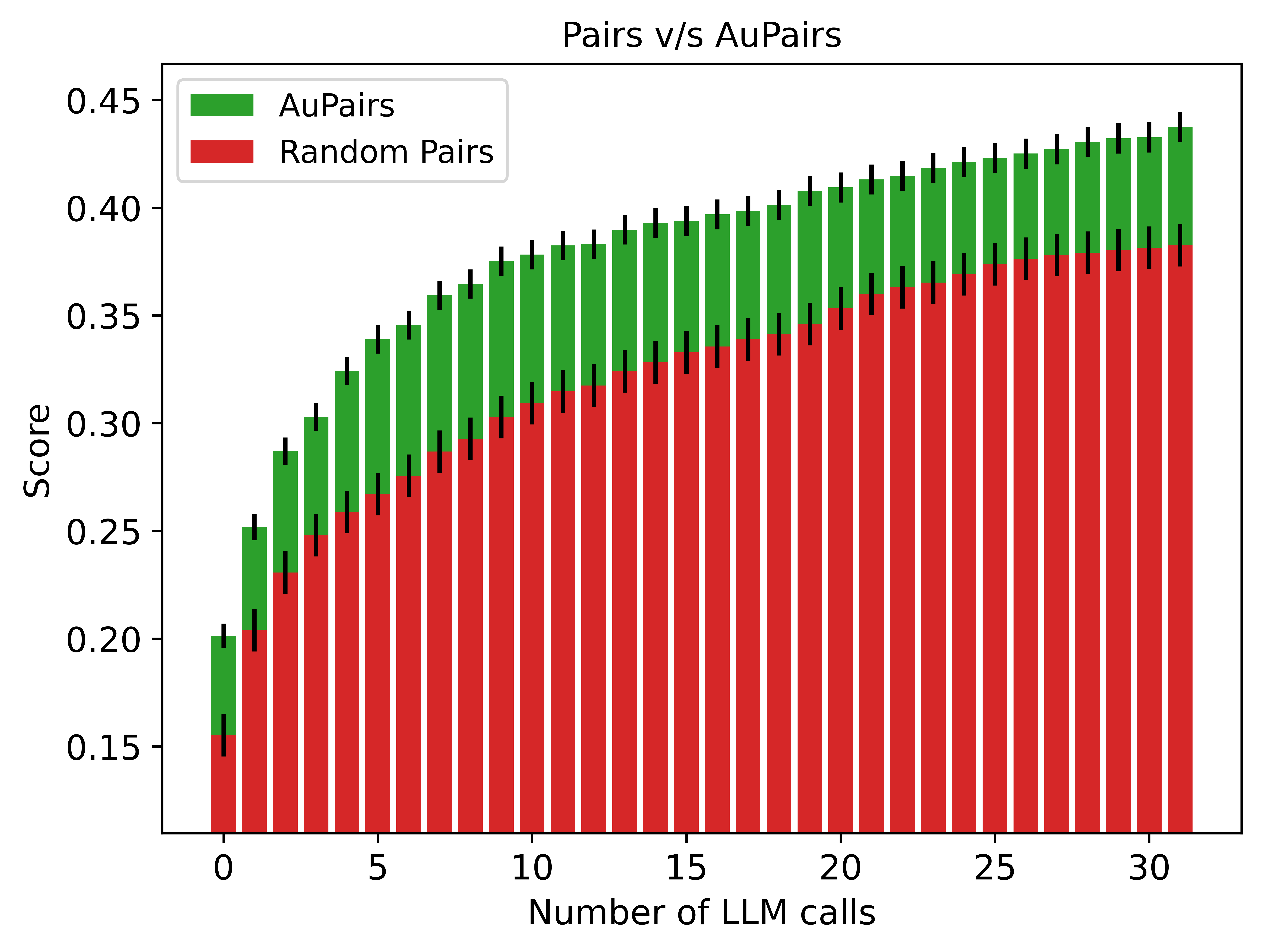}
    \includegraphics[width=0.45\linewidth]{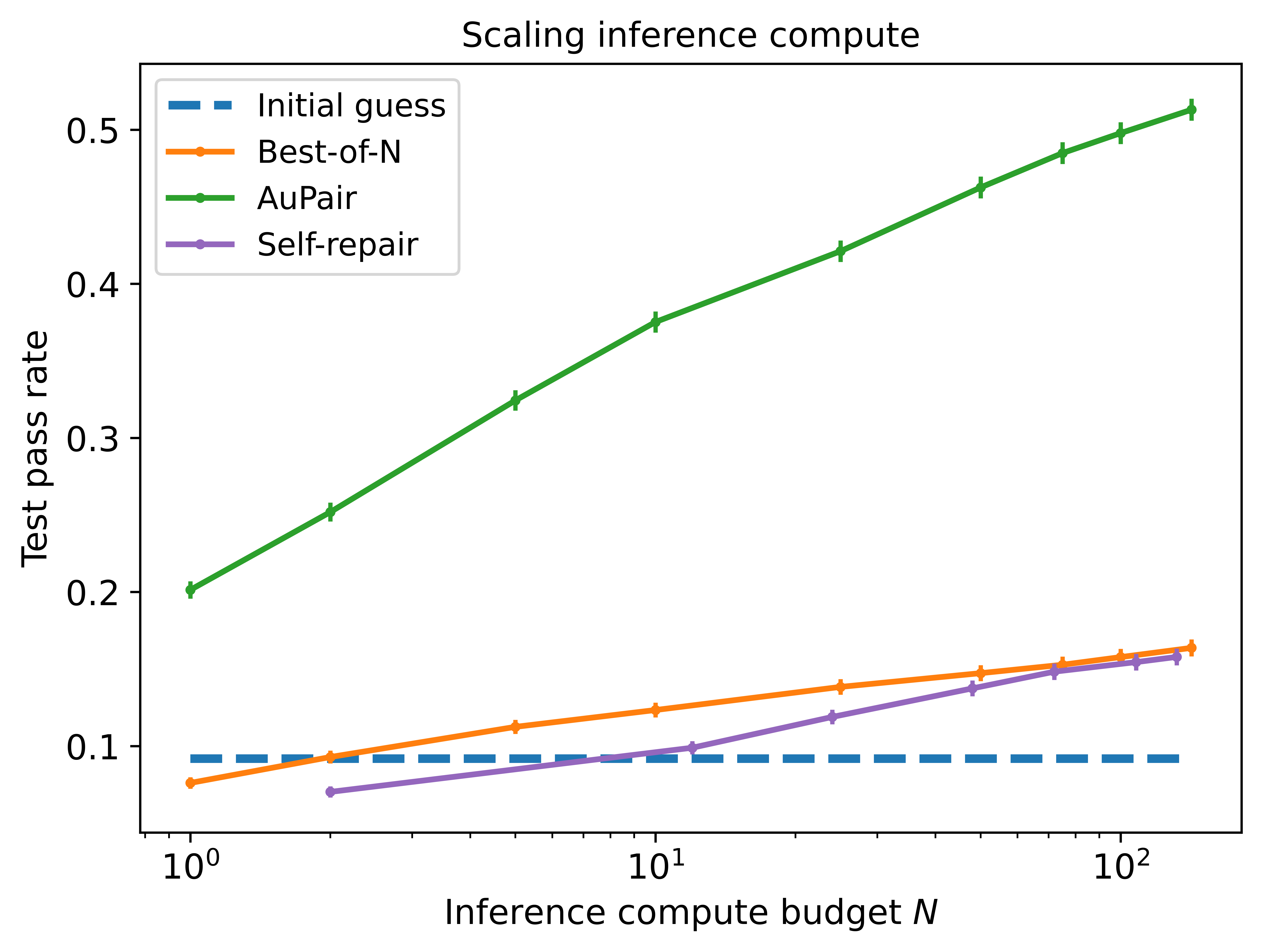}
    \caption{\textbf{(a) \aupairs/ vs. random pairs}: \aupairs/ (green) are significantly (about $2.5-3\times$) more compute-efficient than random pairs (red); it takes only 12 \aupairs/ to reach the same performance as 32 random pairs; \textbf{(b) Scaling inference-time compute:} using \aupairs/ the score increases with compute budget at a much steeper rate compared to baselines (CodeForces dataset, Gemini-1.5-Pro).
    }
    \label{fig:pairs_vs_aupairs}
\end{figure*}

We design an ablation to disentangle the two possible sources of improvement that our approach demonstrates, namely 1) in-context learning and 2) the choice of \aupairs/. It is not implausible for the boost in performance to result from the LLMs' in-context learning ability, and that the same result could be achieved by including \emph{any} set of pairs. On the other hand, our approach specifically targets complementarity during construction of \aupairs/ in that subsequent \aupairs/ are selected based on their ability to solve problems that previous \aupairs/ were unable to solve. To resolve this, we compare the full method to a random-pair baseline that randomly selects pairs from the full candidate set (the result of Phase 1), deduplicating the problems that the random pairs solve (which makes it a stronger baseline). Fig.~\ref{fig:pairs_vs_aupairs} shows that \aupair/ significantly outperforms the random-pair baseline for $N = 1, ..., 32$, saving $2.5-3\times$ more compute by achieving the same score with 12 \aupairs/ as the random pair baseline gets with 32 \aupairs/. Note that for any fixed candidate set, as $N$ grows toward the size of the full set of pairs, the performance of the random-pair baseline will equal that of \aupair/.

\subsection{Better Scaling with Inference-Time Compute}
\label{sec:inference_compute_scaling}

At a fixed budget of $N = 32$ LLM calls, our results look promising. In this section, we investigate whether and how performance scales with the compute budget $N$. Fig.~\ref{fig:pairs_vs_aupairs}(b) plots the score as a function of $N$ using Gemini-1.5-Pro on the CodeForces dataset (additional scaling curves for Gemini-1.5-Flash and strict accuracy metric in the Appendix, see Figs.~\ref{fig:inference_compute_scaling_flash} and \ref{fig:inference_compute_scaling_strict_accuracy}). For each additional LLM call, we use the next best \aupair/ produced by the algorithm and provide it in context to generate the LLM response. Our algorithm produces 144 \aupairs/ for the CodeForces dataset using Gemini-1.5-Pro, and achieves a test pass rate of 51.32\% and strict accuracy of 39.73\% (see \S\ref{sec:inference_compute_scaling_flash}) at 144 LLM calls. The results shows a clear scaling trend with a consistent log-linear performance increase as a function of compute, without any sign of a plateau. More importantly, the increase is substantially \emph{steeper} than the best-of-$N$ and self-repair baselines (which achieve test pass rate of 16.05\% and 15.79\% and strict accuracy 12.04\% and 12.23\% respectively); in other words, prompting with in-context complementary \aupairs/ makes more efficient use of compute than either repeated sampling given a fixed repair prompt, or repair with model-generated verbal feedback.

\subsection{Strong Generalisation to Out-of-distribution Datasets}
\label{sec:OOD}

\begin{figure}[t!]
    \centering
        \includegraphics[width=\linewidth]{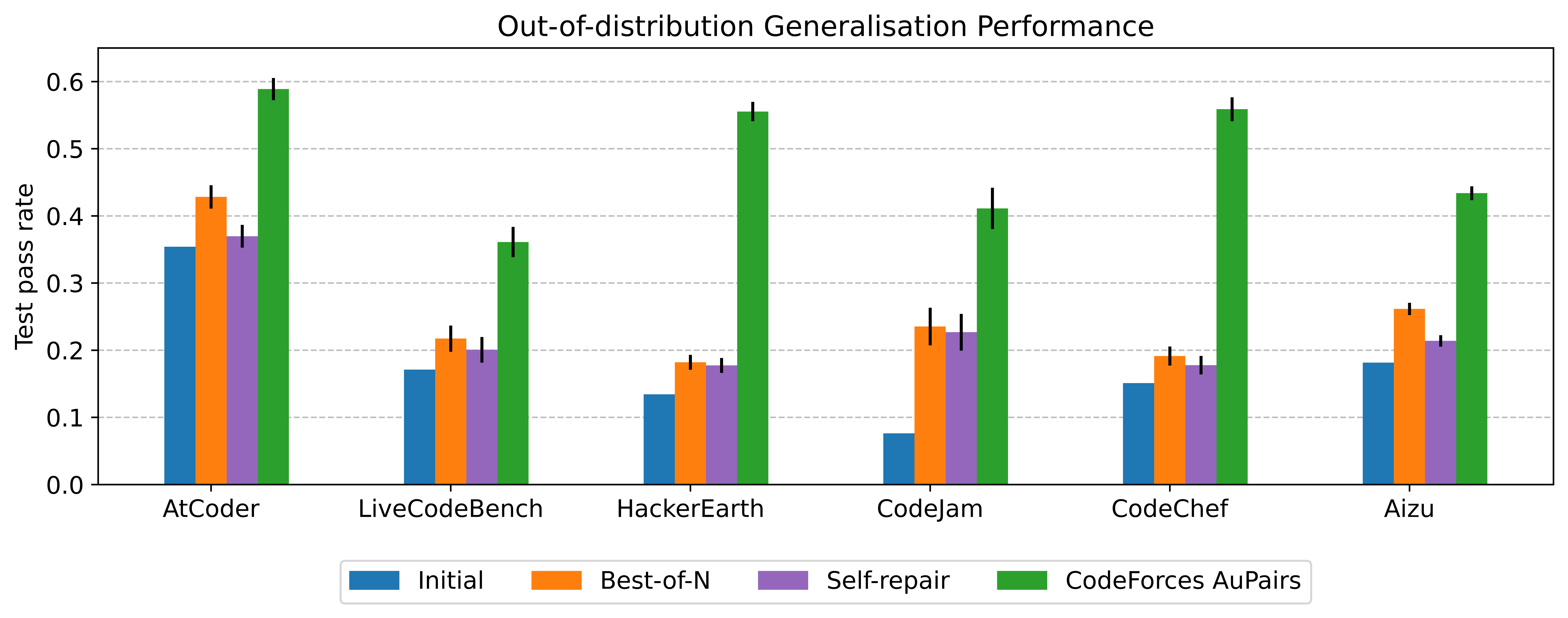}
        \caption{\textbf{Out-of-distribution code repair performance:} \aupairs/ extracted on the CodeForces dataset show strong generalisation performance across the other six datasets (the above results are obtained using Gemini-1.5-Pro and report the test pass rate metric)}
        \label{fig:ood_generalisation}
\end{figure}

The aim of this set of experiments is to determine whether our approach exhibits out-of-distribution generalisation, i.e., given \aupairs/ collected on a different dataset, see if we can retain the performance improvements that we obtain in-distribution. To test out-of-distribution generalisation, we evaluate the \aupairs/ collected using the Gemini-1.5-Pro model on the CodeForces dataset on the other 6 datasets and compare them with the corresponding baselines. Fig.~\ref{fig:ood_generalisation} shows that for all 6 datasets, our approach outperforms both baselines by a large margin, in spite of having out-of-distribution \aupairs/. This in turn implies that the process of collecting \aupairs/ may only be needed on one dataset, and its benefits can be reaped across a wide range of problems (from other datasets, or users) at inference time. We also observe matching performance for in-distribution AtCoder \aupairs/ and out-of-distribution CodeForces \aupairs/ on the AtCoder dataset.

\subsection{Decent Cross-Model Transfer}
\label{sec:model_transfer_analysis}

\begin{figure}[t]
    \centering
    \begin{minipage}{0.5\linewidth}
        \includegraphics[width=\linewidth]{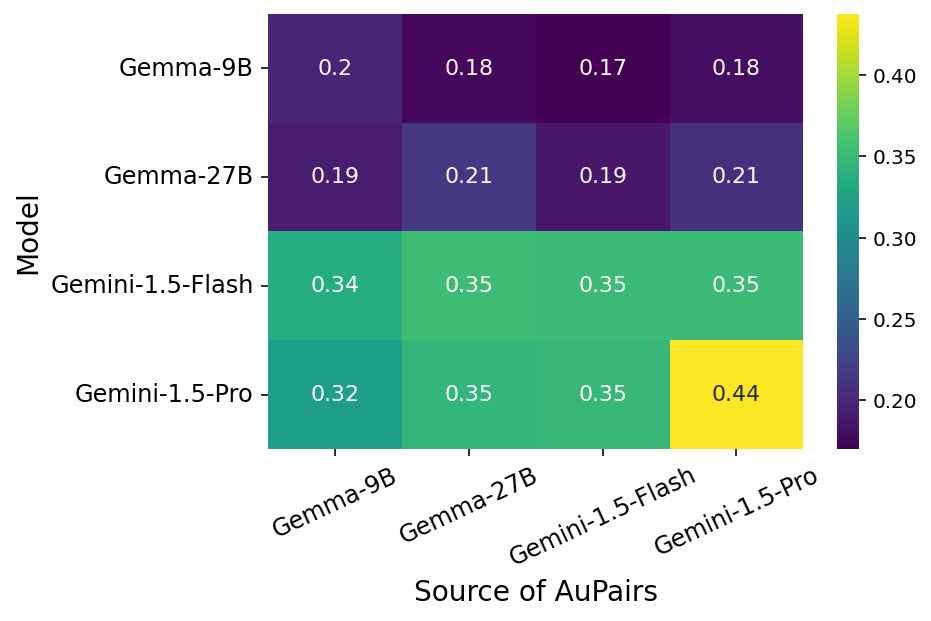}
    \end{minipage}
    \begin{minipage}{0.48\linewidth}
        \includegraphics[width=0.87\linewidth]{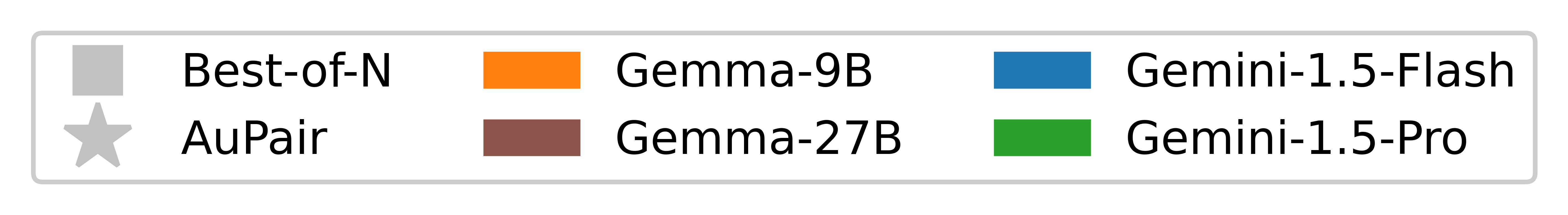}\\
        \centering
        \includegraphics[width=0.80\linewidth]{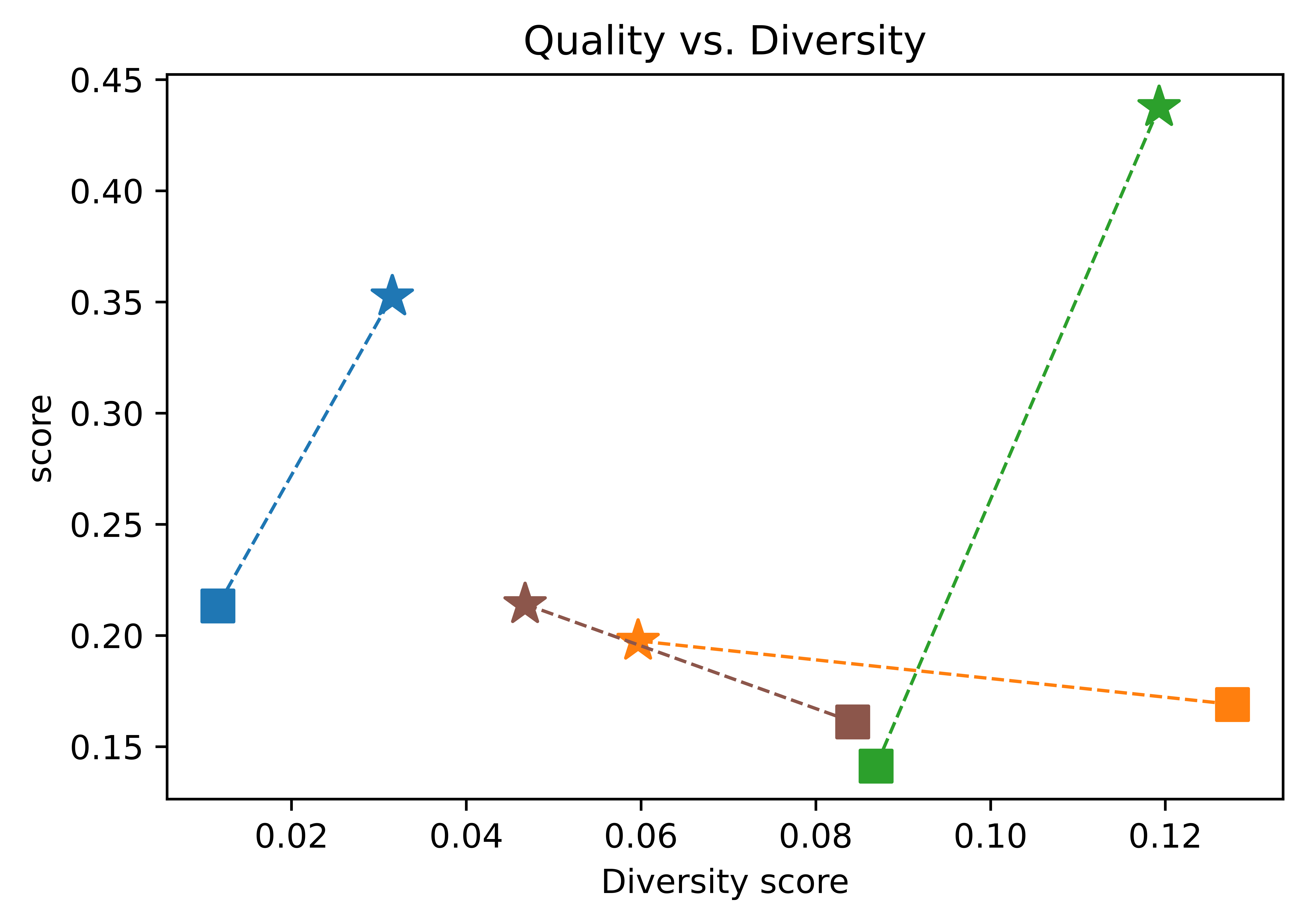}
    \end{minipage}
    \caption{\textbf{(a) Cross-model transfer:} \aupair/ shows good cross-model transfer capabilities for several model combinations on CodeForces; \textbf{(b) Diversity-Score plot:} \aupair/ ($\star$) with Gemini-1.5-Flash and Gemini-1.5-Pro generates more diverse responses than best-of-$N$ ($\Box$) while this diversity trend is reversed for the Gemma models. In terms of score, \aupair/ always generates higher-scoring fixes than best-of-$N$. Details on diversity computation are given in \S\ref{sec:aupair_analysis_diversity}.}
    \label{fig:cross_model_transfer}
\end{figure}

Now that we have seen that our approach can exhibit very good out-of-distribution generalisation along the data axis, we evaluate it on its ability to generalise on the model axis, i.e., we look at the performance of \aupairs/ collected using a different model. We evaluate this cross-model transfer capability for several model combinations on CodeForces. The resulting 16 ablations are shown in Fig.~\ref{fig:cross_model_transfer}(a), and help disentangle the impact of the \aupairs/ versus the code repair capabilities of the inference model. A key takeaway is that the Gemma models exhibit worse performance, regardless of the quality of \aupairs/ used at inference time, indicating that they are inferior at the capability of code repair. Gemini-1.5-Flash performs much better at code repair, and its sensitivity to the source of \aupairs/ is negligible: it is equally performant for each source. Gemini-1.5-Pro, on the other hand, \emph{is} sensitive to the source of \aupairs/; in particular, when Gemini-1.5-Pro uses \aupairs/ collected by the same model, it achieves the best performance by a large margin. With \aupairs/ selected using other models, Gemini-1.5-Pro achieves comparable performance to Gemini-1.5-Flash. One reason for the standout performance when using Gemini-1.5-Pro \aupairs/ seems that those examples result in substantially more diverse generations, as shown in  \S\ref{sec:aupair_analysis_diversity}. However, Fig.~\ref{fig:cross_model_transfer}(a) as a whole suggests that there is an ordering in terms of performance: 1) the model used at inference time has to have good code repair capabilities, and 2) the stronger the model is at code repair, the more improvement we can expect from it with a higher \emph{quality} of \aupairs/.

\subsection{High Code-specific Diversity} 
\label{sec:aupair_analysis_diversity}

We dive a bit deeper into the nature of fixes generated using different \aupairs/. There are several ways to analyse code; we choose Abstract Syntax Trees (ASTs) since they mostly capture the structure of changes. More concretely, since we have $N$ fixes for each problem (here $N = 32$), we measure the diversity per problem as the number of unique changes made to the guess over all $N$ fixes for that problem. The diversity score is calculated as the average number of unique abstract syntactic subtrees generated per problem. More concretely, we perform the set difference of all subtrees in the fix AST that are not in the guess AST and normalise with the maximum number of subtrees.  We plot this diversity metric against the score in Fig.~\ref{fig:cross_model_transfer}(b) to get a sense of how diverse and useful the \aupairs/ are. We also include diversity results of the best-of-$N$ baseline, see \ref{sec:appendix_diversity} for further details on the diversity score computation. The results show that while \aupairs/ always increase performance, they result in higher diversity of fixes when given to the more competent models (Gemini-1.5-Pro and -Flash), and lower diversity for Gemma models. It is worth highlighting that the exceptional performance of \aupairs/ produced and used by Gemini-Pro (Fig.~\ref{fig:cross_model_transfer}(a), bottom right) corresponds to highly diverse fixes (Fig.~\ref{fig:cross_model_transfer}(a), top right).

\begin{table}[t]
\centering
\scalebox{0.8}{
\begin{tabular}{lcccccc}
\toprule
 Difficulty level $\rightarrow$ & A (671)  & B (675) & C (671)& D (666) & E (649)& F+ (537) \\
 \midrule
Gemma-9B&\cellcolor{green1}0.34 {\small(+0.16)} &\cellcolor{green0}0.23 \small(+0.13)&\cellcolor{green0}0.19 \small(+0.12)&\cellcolor{green0}0.15 \small(+0.09)&\cellcolor{green0}0.14 \small(+0.08)&\cellcolor{green0}0.12 \small(+0.07)\\
Gemma-27B&\cellcolor{green1}0.28 \small(+0.1)&\cellcolor{green1}0.25 \small(+0.12)&\cellcolor{green1}0.20 \small(+0.12)&\cellcolor{green0}0.19 \small(+0.1)& \cellcolor{green0}0.17 \small(+0.1)&\cellcolor{green1}0.20 \small(+0.11)\\
Gemini-1.5-Flash&\cellcolor{green3}0.54 \small(+0.2)&\cellcolor{green2}0.39 \small(+0.18)&\cellcolor{green2}0.34 \small(+0.15)&\cellcolor{green0}0.18 \small(+0.11)&\cellcolor{green1}0.26 \small(+0.12)&\cellcolor{green1}0.28 \small(+0.11)\\
Gemini-1.5-Pro&\cellcolor{green3}0.62 \small(+0.42)&\cellcolor{green3}0.52 \small(+0.4)&\cellcolor{green2}0.43 \small(+0.35)&\cellcolor{green2}0.38 \small(+0.32)&\cellcolor{green1}0.32 \small(+0.28)&\cellcolor{green1}0.35 \small(+0.29)\\
\bottomrule
\end{tabular}
}
\caption{\textbf{Difficulty-wise analysis:} test pass rate using \aupairs/, categorised by difficulty level from easy (A) to hard (F+), accompanied by number of problems. Absolute improvement in parentheses. We see an expected trend here: the strongest performance is observed using the best models on the easiest problems, and as difficulty increases, performance decreases across models. However, our results with Gemini-1.5-Pro indicate improved performance with higher difficulty}
\label{tab:difficulty}
\end{table}

\subsection{Improvement on All Difficulty Levels}

Coding datasets have heterogeneous difficulty. As a sanity check, we conduct additional analysis to determine \emph{which problem levels} are most helped by \aupair/, compared to the quality of initial guesses. Table~\ref{tab:difficulty} shows the absolute improvement in test pass rate, i.e., the increase in this score achieved by \aupair/ for 4 models on CodeForces. The two key observations are (a) \aupair/ helps significantly at all difficulty levels across models, and (b) there are larger improvements on easier levels, and this trend is consistent across models. Note that the initial performance of Gemini-1.5-Pro is low because the initial guesses generated do not adhere to the instruction (elaborated in Appendix~\S\ref{sec:code_execution}); however since this is the strongest model and shows the best overall performance across difficulty levels, the increases in score that we see are significantly higher than the other models.

\subsection{Coverage of Problem Categories is Preserved}
\label{sec:categories}

\begin{figure}[t]
    \centering
    \includegraphics[width=0.99\linewidth]{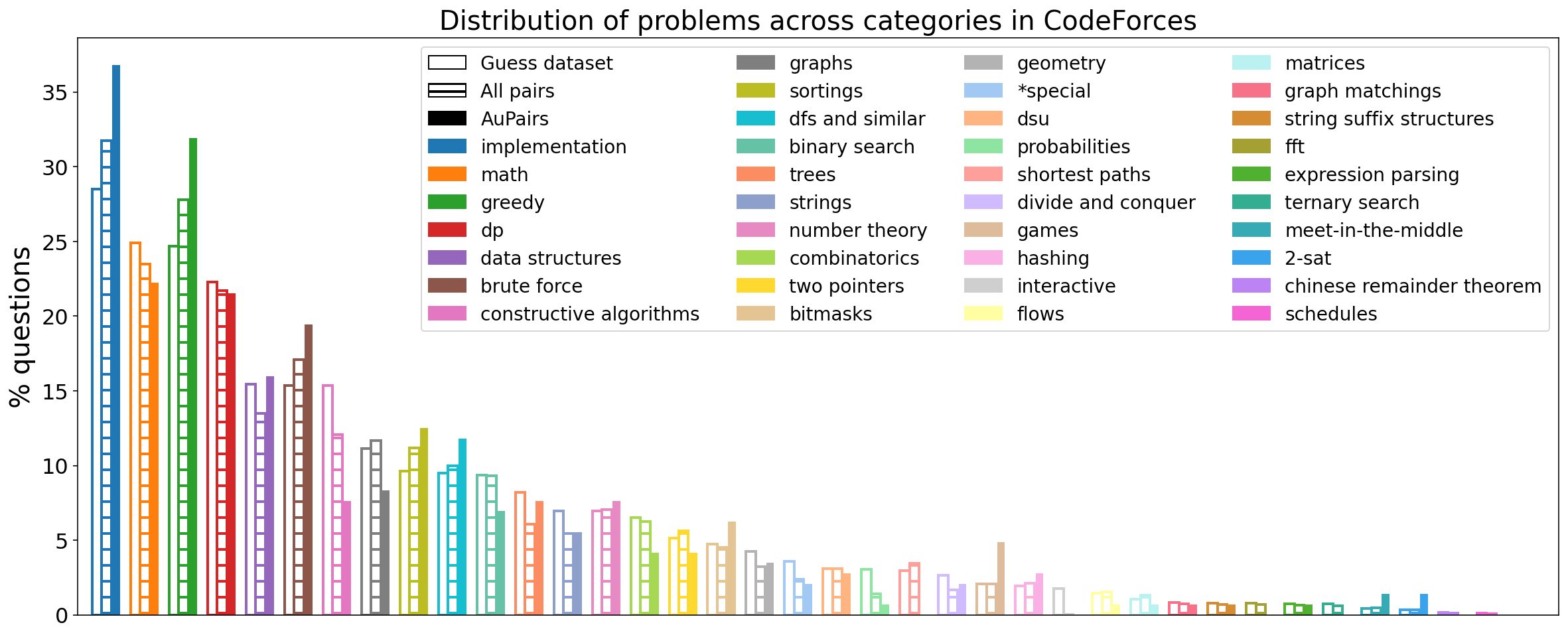}
    \caption{\textbf{Category-wise analysis:} analysing the distribution of \aupairs/ across different categories and comparing it with the distribution of problems in the dataset.}
    \label{fig:category_analysis}
\end{figure}

The CodeForces dataset is richly annotated with category labels for each problem. A problem may have multiple tags, for instance, $\texttt{\small strings}$ and $\texttt{\small two pointers}$. We use these fine-grained tags to study how the problem distribution is affected by Phase 1 and Phase 2 of our method, separately. Fig.~\ref{fig:category_analysis} shows the proportions of these categories observed in the initial dataset, the full set of pairs generated during Phase 1, and the final \aupairs/. The high-level result is encouraging, namely that the starting diversity is approximately preserved. Phase 1 yields pairs for every single category, even those that lie at the tail. Furthermore, the (sparser) distribution over categories for the \aupairs/ after Phase 2 still shows several problems from rare categories. This additional result consolidates our insight that \aupairs/ are highly diverse, also in the types of problems they contain.

\section{Related Work}
\label{sec:related_work}
Automatic Program Repair (APR) has been a longstanding research area in the field of machine learning~\citep{devlin2017semcode, bhatia2016synfix, chen2019sequencer, feng-etal-2020-codebert, berabi2021tfix, codit, circle}. Most methodologies rely on supervised finetuning to adapt LLMs to the task of code generation using labeled pairs of broken / fixed code pairs, which is costly to obtain and often task- and problem-specific~\citep{Hu2022nsedit,cure,xia2022alpharepair, dinella2020hoppity}. On the other hand, unsupervised APR is challenging since it requires syntactic and semantic understanding of code, and most automatic code breaking approaches tend to be out-of distribution with real samples. \citet{yasunaga2021pmlr} train both a breaker and a fixer in order to learn to propose new code fixes that are realistic, and use a compiler to verify its correctness. Our work uses partial fixes generated by the model as the initial broken code to be fixed iteratively.

More recently, a few unsupervised approaches have been proposed based on the capability of LLMs to generate code~\citep{chen2021evaluating,nijkamp2023codegen, chowdhery2024palmcoder, li2022alphacode, fried2023incoder, li2023starcoder}. The APR task still remains challenging, even though models are better at generating code~\citep{olausson2024self,chen2023teaching}. \citet{zhao2024repair}~use a step-by-step method to repair code using a reward model as a critic, providing feedback to finetune an LLM. \citet{shypula2024codeedits} propose a retrieval based few-shot prompting approach with Chain-of-Thought (CoT) reasoning traces, and use supervised fine-tuning (SFT) to finetune a model using self-play.

The main disadvantage of using SFT approaches comes from the need to finetune the model to the task, which becomes much more costly with ever-growing model sizes. In recent years the in-context learning (ICL) paradigm~\citep{brown2020icl} has been shown to be a flexible and compute-efficient adaptation approach to new tasks~\citep{oswald2022iclgrad, akyurek2023ICLlinear}. \citet{le2022coderl} use an LLM to generate code and a critic network to predict functional correctness of the the generated program, with zero-shot transfer to new tasks. Our work focuses on tasks in which the correctness is measured by the number of test cases the generated code passes. \citet{Gou2024critic} combine the use of LLMs with tools to provide feedback for the LLM to self-correct via additional calls to evaluate its own output in a validation setting. \citet{wang2024intervenor} also make use of external tools and use an LLM in a learner / teacher role to provide a chain of repairs to fix the code.

\citet{xin2024thinkrepair} propose an automated self-repair approach with few-shot prompting but using CoT and execution feedback information. \citet{agarwal2024manyshot} also use CoT rationales but remove them from context when few-shot-prompting the model. \citet{olausson2024self} show that using an LLM as a feedback source for self repair has its limitations when compared with the same number of independent model calls for the same problem since the ability to generated better code may be interconnected with the ability to identify its faulty behaviour. \citet{welleck2023selfcorrect} decouple the generation and correction phases by independently training a corrector with scalar and natural language feedback to correct intermediate imperfect generations. We use self-corrections, since we use the same model for generating the fixes and the broken code pairs, but the improvement is grounded on the number of passing tests, avoiding degenerate behaviours.

\citet{yuan2017arja} propose a multi-objective evolutionary algorithm to search over possible correct code patches; \citet{paredes2023funsearch} use an island-based evolutionary method to encourage exploration of diverse programs, and perform iterative best-shot-prompting to improve the quality of the generated code. We use a generative approach; closer to the work of~\citet{shirafuji2023fewshot}, we make use of ICL abilities of LLMs to generate improved code repairs, but we provide an extra submodular process to select the samples, that encourages diversity.

\section{Conclusions and Future Work}
\label{sec:conclusions}
We propose a novel algorithm, which produces an ordered set of \aupairs/, each of which can be provided as an in-context example using 1-shot prompting with an inference compute budget of $N$ LLM calls to improve code repair performance at inference time. Our approach is highly scalable, showing significantly better outcomes than best-of-$N$ and self-repair, both of which are known to improve performance as inference compute is scaled up. In addition to this, the \aupairs/ generated using our algorithm show strong out-of-distribution generalisation and thus can be reused at inference time to solve a wide range of problems. While in this paper we have explored repair in the coding domain, future work can look at using it in other settings in which an initial solution generated by an LLM can be improved via repair. Additionally, the choice of coding implies that all our feedback is grounded, but using ungrounded feedback from reward models to build the \aupairs/ might be another potential direction worth exploring.

\section{Acknowledgements}

The authors would like to thank Dan Calian for valuable feedback and suggestions on the paper; David Silver for his sponsorship; Wojtek Czarnecki, Kate Baumli, Jakub Sygnowski, Victor C\u{a}rbune, Volodymyr Mnih, Mina Khan, Georg Ostrovski, Shantanu Thakoor, Lei Zhang, Disha Shrivastava, Feryal Behbahani, the RL team, and the wider DeepMind community for helpful discussions.

\section{Author Contributions}

\textbf{Aditi Mavalankar:} \aupair/ concept, algorithm design, project leadership, infrastructure, experimentation, analysis, paper writing

\noindent\textbf{Hassan Mansoor:} infrastructure, data curation, experimentation

\noindent\textbf{Zita Marinho:} prompting, analysis, paper writing

\noindent\textbf{Masha Samsikova:} prompting, initial infrastructure

\noindent\textbf{Tom Schaul:} algorithm design, project leadership, paper writing, sponsorship

\typeout{}
\bibliography{main}

\clearpage

\appendix
\section{Appendix}
\label{sec:appendix}

\subsection{Pair Generation}

In this section, we discuss the specifics of the pair generation phase and provide results pertaining to this phase. The approach that we use for pair generation is provided in Algorithm~\ref{alg:pair_generation}. Note that this is one way to generate pairs; they can be generated in other ways, or be available beforehand. Studying the impact of using pre-generated pairs for extracting \aupairs/ could be an interesting avenue for future work.

\begin{algorithm}[H]
\begin{algorithmic}[1]
\Require
$\left\{\begin{array}{ll}
\text{LLM} & \text{large language model} \\
\mathcal{D_\text{train}} & \text{training dataset}\\
k & \text{number of few-shot examples} \\
N & \text{total number of LLM calls} \\
\text{score} & \text{code eval function} \\
\end{array}
\right.$
\State init candidate pairs $\mathcal{C} \gets \{\}$
\For{$i = 1, \ldots, N$}
\State sample problem from dataset: $\bm{x} \sim \mathcal{D}_{\text{train}}$
\State sample $k$ pairs to use in-context: $\bm{c}_1, \ldots, \bm{c}_k \sim \mathcal{C}$
\State build $k$-shot prompt: $\bm{p} \gets \bm{c}_1 \mathbin\Vert \ldots \mathbin\Vert \bm{c}_k \mathbin\Vert \bm{x}$
\State generate fix: $\bm{\hat{y}} \gets \text{LLM}(\bm{p})$
\State evaluate fix: $s_{\bm{\hat{y}}} \gets \text{score}(\bm{\hat{y}})$
\If {$s_{\bm{\hat{y}}} > s_{\bm{x}}$}
    \State create new pair: $\bm{c} \gets \langle \bm{x}, \bm{\hat{y}} \rangle$
    \State add to candidate pairs: $\mathcal{C} \gets \mathcal{C} \cup \bm{c}$
    \If {$s_{\bm{\hat{y}}} < 1$}
        \State create new problem $\bm{\hat{x}}$ with guess $\bm{\hat{y}}$
        \State add new problem to dataset: $\mathcal{D}_\text{train} \gets \mathcal{D}_\text{train} \cup \bm{\hat{x}}$
    \Else
        \State remove all instances of problem from dataset: $\mathcal{D}_\text{train} \gets \mathcal{D}_\text{train} - \{ \bm{x} \}$
    \EndIf
\EndIf
\EndFor
\Statex \Return{$\mathcal{C}$}
\end{algorithmic}
\caption{Pair Generation}
\label{alg:pair_generation}
\end{algorithm}

We set $k = 32$ since during pair generation, we want diverse pairs to be generated, and using a different set of $k$ examples with the same problem could give us different fixes.

For the AtCoder dataset, we set a budget of 10,000 LLM calls for pair generation. Since the CodeForces dataset is larger, we set a budget of 35,000 LLM calls to maintain a good balance between having enough LLM calls per problem and maintaining the affordability of the overall approach in terms of computational resources. We report the number of pairs generated on both of these datasets across all 5 models: Gemini-1.5-Pro, GPT-4o-mini, Gemini-1.5-Flash, Gemma-27B, and Gemma-9B in Table~\ref{tab:number_of_pairs}. Here we provide some additional results that we were unable to include in the main text.

\begin{table}[H]
    \centering
    \begin{minipage}{0.45\linewidth}
        \centering
        \begin{tabular}{lcc}
            \toprule
             CodeForces & \# Pairs & \# \aupairs/\\
             \midrule
             Gemini-1.5-Pro & 1560 & 144\\
             \rowcolor{lavender}
             GPT-4o-mini & 1192 & 94\\
             Gemini-1.5-Flash & 1327 &110\\
             \rowcolor{lavender}
             Gemma-27B & 509 & 77\\
             Gemma-9B & 556 & 122\\
            \bottomrule
        \end{tabular}
    \end{minipage}
    \begin{minipage}{0.45\linewidth}
        \centering
        \begin{tabular}{lcc}
            \toprule
             AtCoder & \# Pairs & \# \aupairs/ \\
             \midrule
             \rowcolor{lavender}
             Gemini-1.5-Pro & 927 & 64\\
             GPT-4o-mini & 378 & 15\\
             \rowcolor{lavender}
             Gemini-1.5-Flash & 397 & 64\\
             Gemma-27B & 295 & 27\\
             \rowcolor{lavender}
             Gemma-9B & 147 & 14\\
            \bottomrule
        \end{tabular}
    \end{minipage}

\caption{Number of pairs collected during phase 1 of the algorithm ($\#$ of pairs) and number of \aupairs/  extracted in  phase 2 ($\#$ \aupairs/): CodeForces (left) and AtCoder (right) for all 5 models.}
\label{tab:number_of_pairs}
\end{table}

\subsection{Measuring correctness in terms of solved problems}\label{sec:strict_accuracy}

In addition to pass rate of unit tests, we also report the percentage of fully solved problems, for which the generated code passes all test cases. 

$$\text{strict accuracy} = \frac{1}{P}\sum_{p=1}^{P}\max_{i \in \{1, \ldots, N\}} \prod_{j=1}^{|T_p|} \mathbb{1} \{ \texttt{eval}(\text{code}_{p, i}, T_{p, j}) == \texttt{pass} \}$$

where $T_p$ refers to the unit tests for problem $p$, and $\text{code}_{p, i}$ is the code generated by the LLM for problem $p$ in the $i^{\text{th}}$ LLM call. Here, the innermost loop, like test pass rate, computes the percentage of unit tests passed by the LLM output $\text{code}_{p, i}$. Following this, we select the code output that passes all tests (max over binary values yields 1 if any such output exists, otherwise 0). The outermost loop averages this across all the problems in the test dataset. 

\begin{figure}[t]
    \centering
    \includegraphics[width=0.45\linewidth]{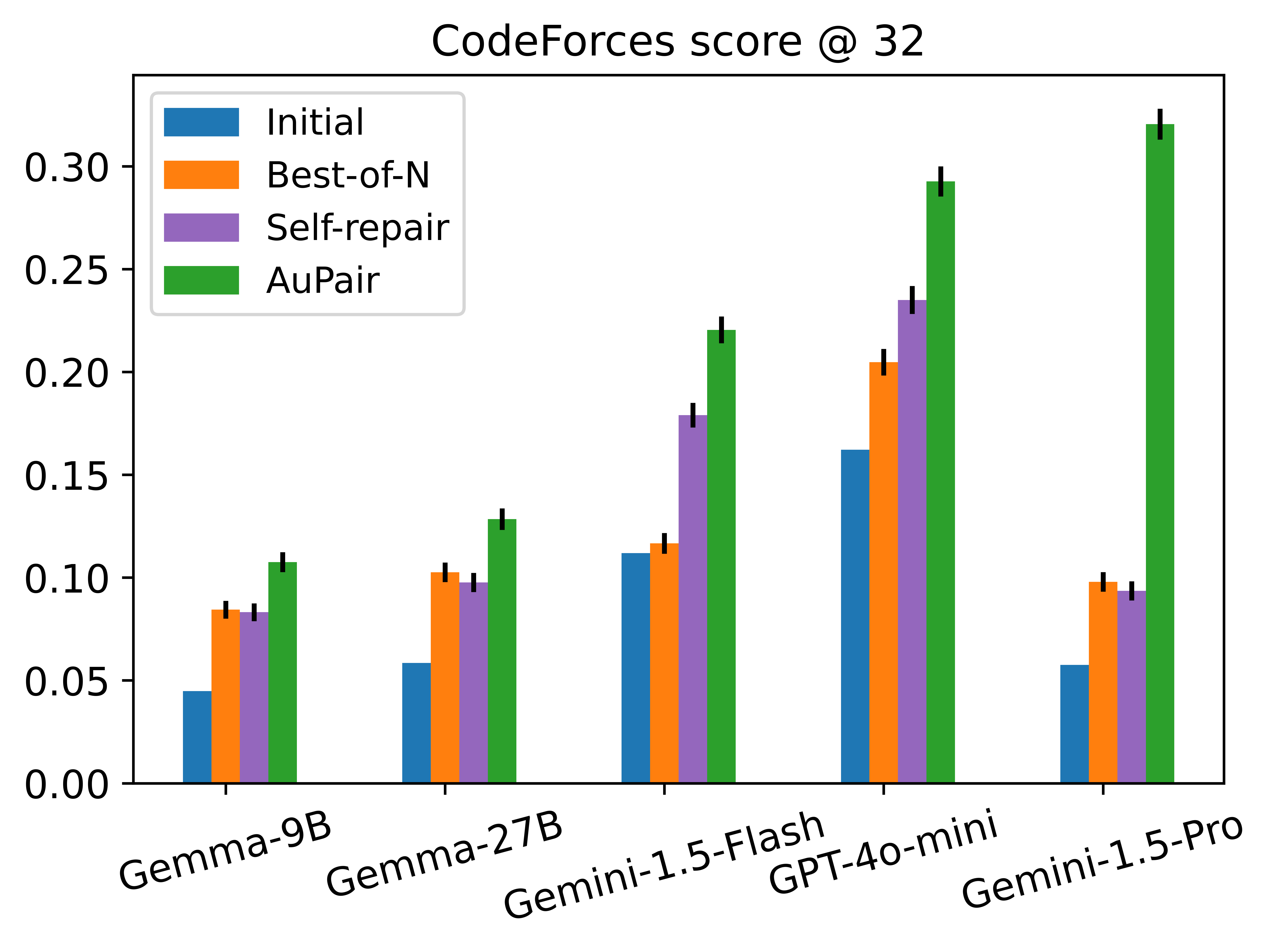}
    \includegraphics[width=0.45\linewidth]{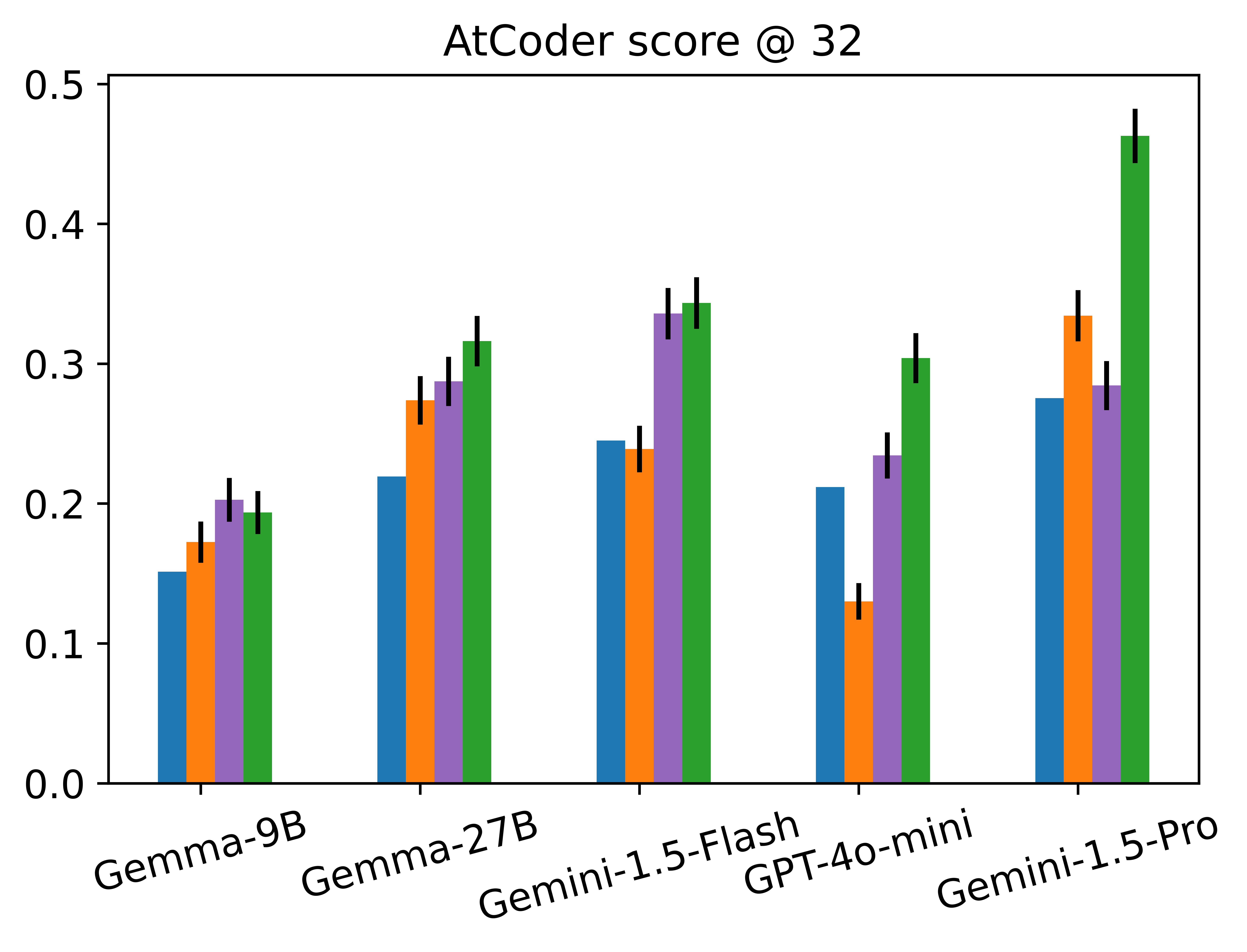}
    \caption{\textbf{In-distribution code repair performance for the strict accuracy metric} with $N = 32$ LLM calls at inference time. CodeForces (left) and AtCoder (right).}
    \label{fig:strict_accuracy_results}
\end{figure}

\begin{figure}[t]
    \centering
        \includegraphics[width=\linewidth]{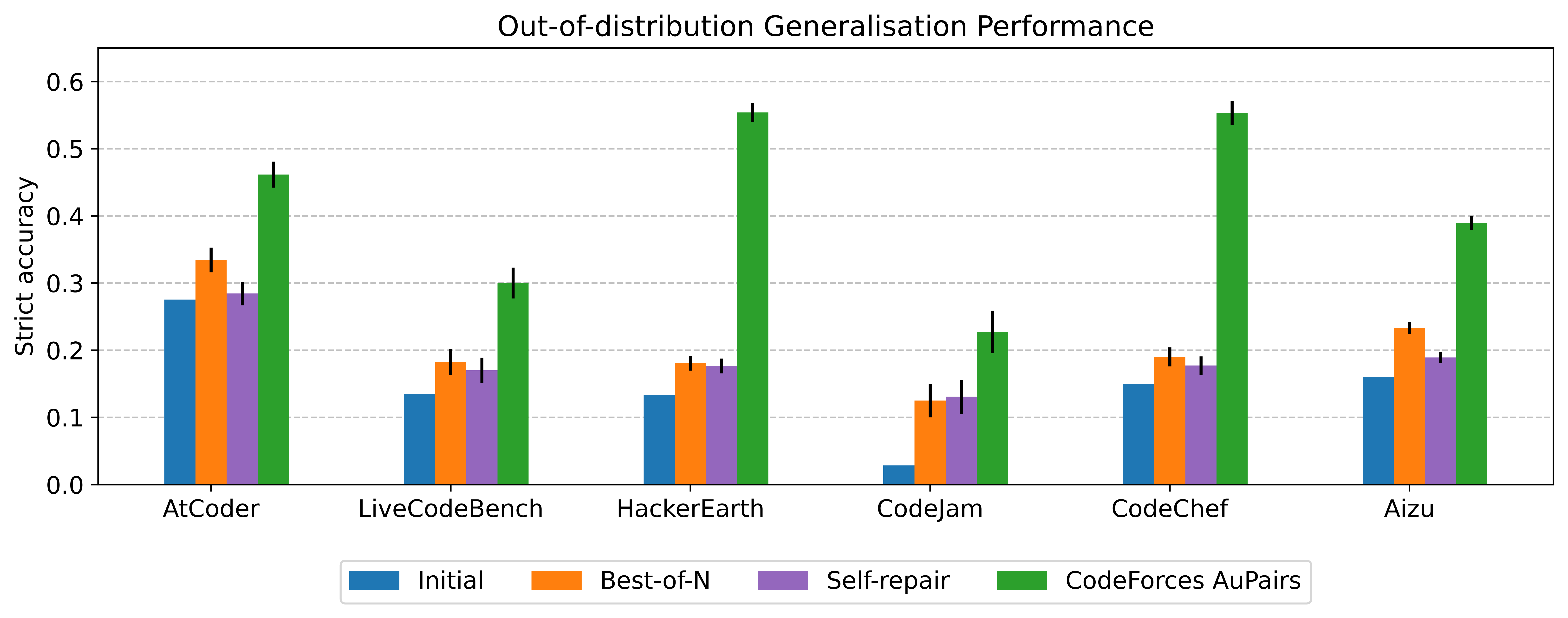}
        \caption{\textbf{Out-of-distribution code repair performance for the strict accuracy metric:} \aupairs/ extracted on the CodeForces dataset show strong generalisation performance across the other six datasets (the above results are obtained using Gemini-1.5-Pro)}
        \label{fig:ood_generalisation_strict_accuracy}
\end{figure}

We see that \aupair/ outperforms all other baselines on all models across the board, with results for CodeForces and AtCoder shown in Fig. \ref{fig:strict_accuracy_results}. We also show the results for out-of-distribution generalisation on this strict accuracy metric in Fig.~\ref{fig:ood_generalisation_strict_accuracy}; again, the results clearly indicate that \aupair/ outperforms all baselines on this metric as well across all datasets. Furthermore, the scaling results in \S\ref{sec:inference_compute_scaling_flash} also indicate that the scaling trends for the strict accuracy metric are consistent with those of the test pass rate metric.

\subsection{Scaling Inference Compute}\label{sec:inference_compute_scaling_flash}

\begin{wrapfigure}{R}{0.5\linewidth}
    \centering
        \includegraphics[width=\linewidth]{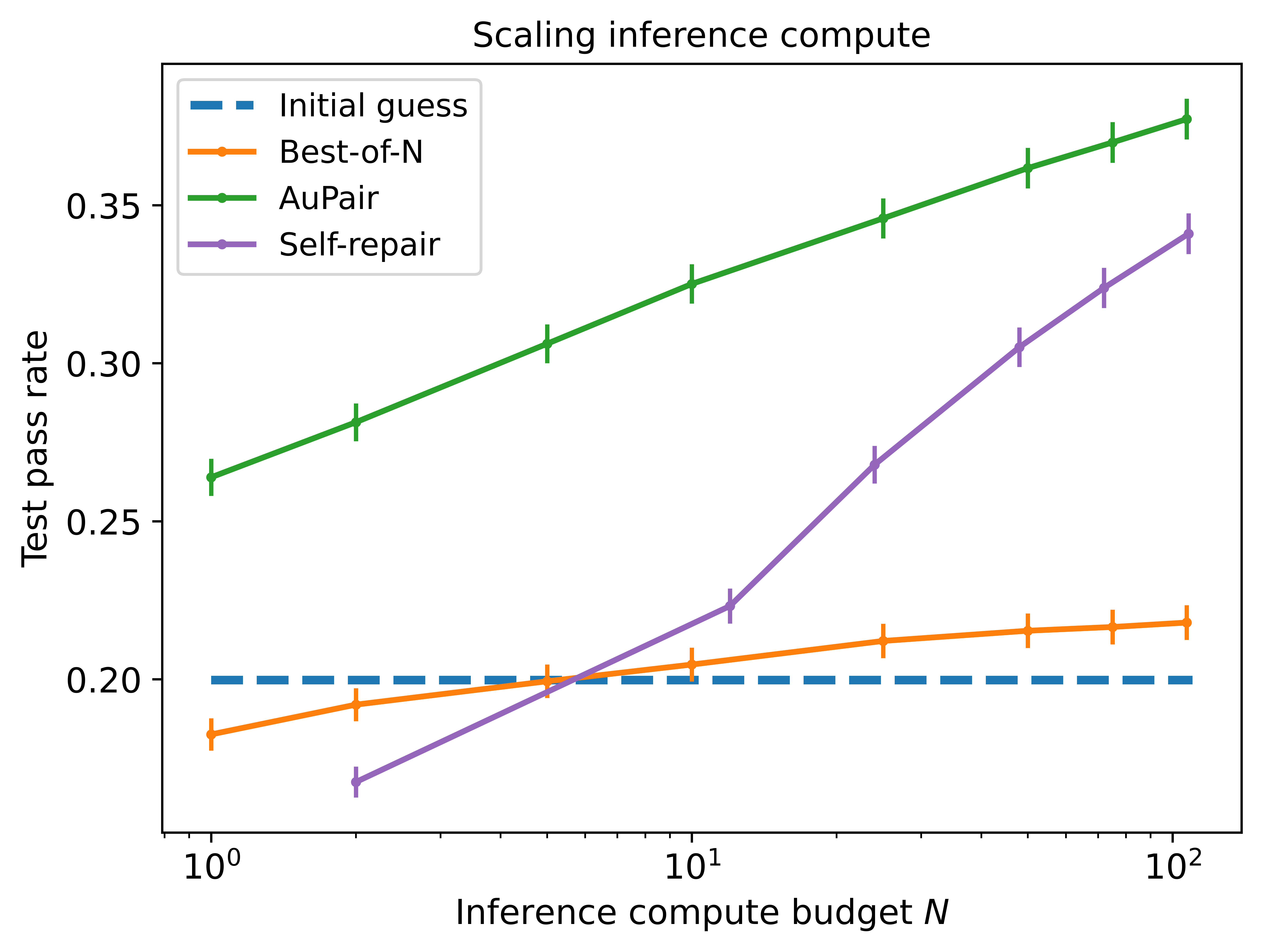}
    \caption{Scaling up inference compute on the CodeForces dataset with Gemini-1.5-Flash. Scores correspond to average pass test rate on all the test problems.}
    \label{fig:inference_compute_scaling_flash}
\end{wrapfigure}

\begin{figure}[t!]
    \centering
    \includegraphics[width=0.45\linewidth]{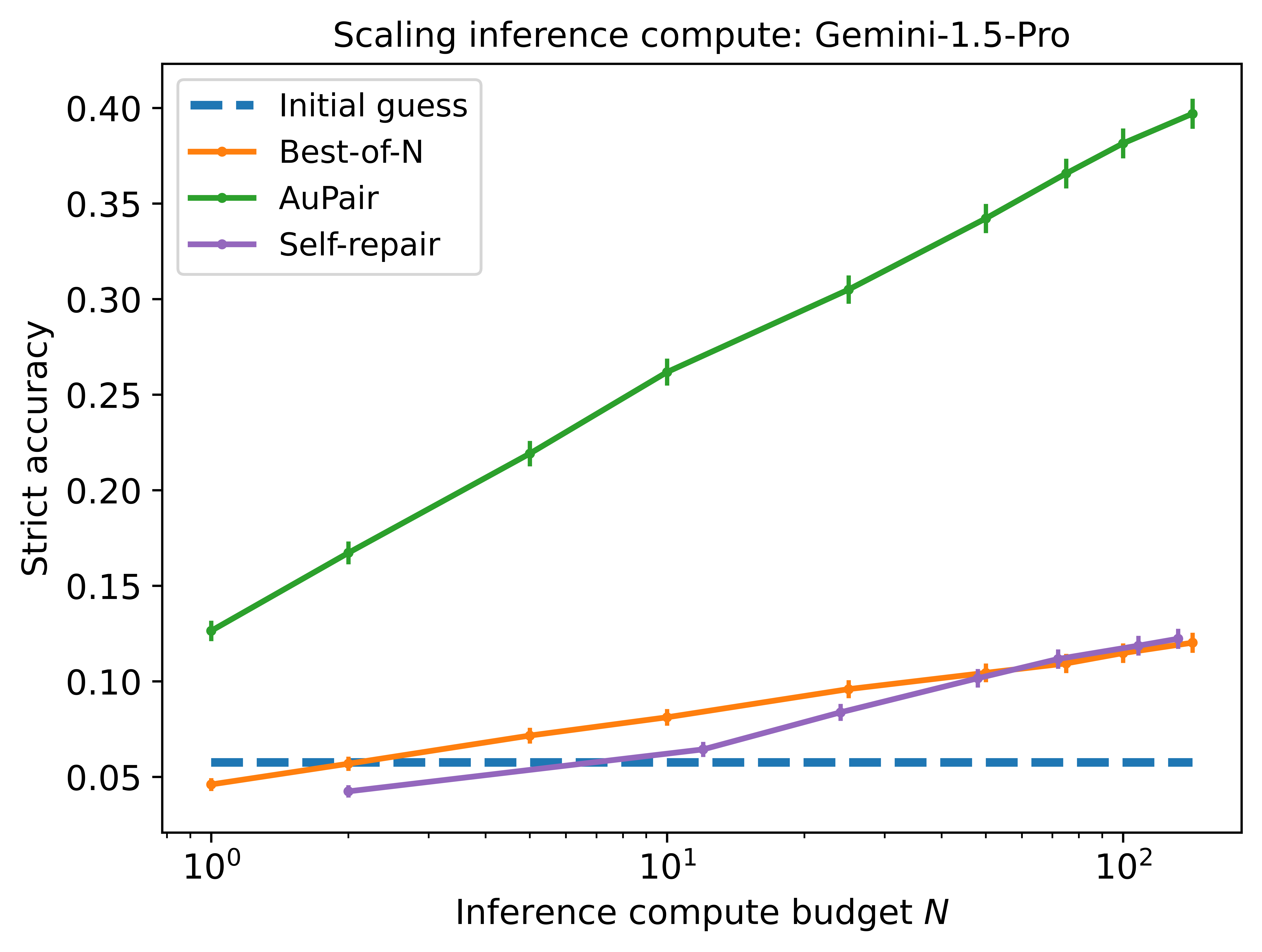}
    \includegraphics[width=0.45\linewidth]{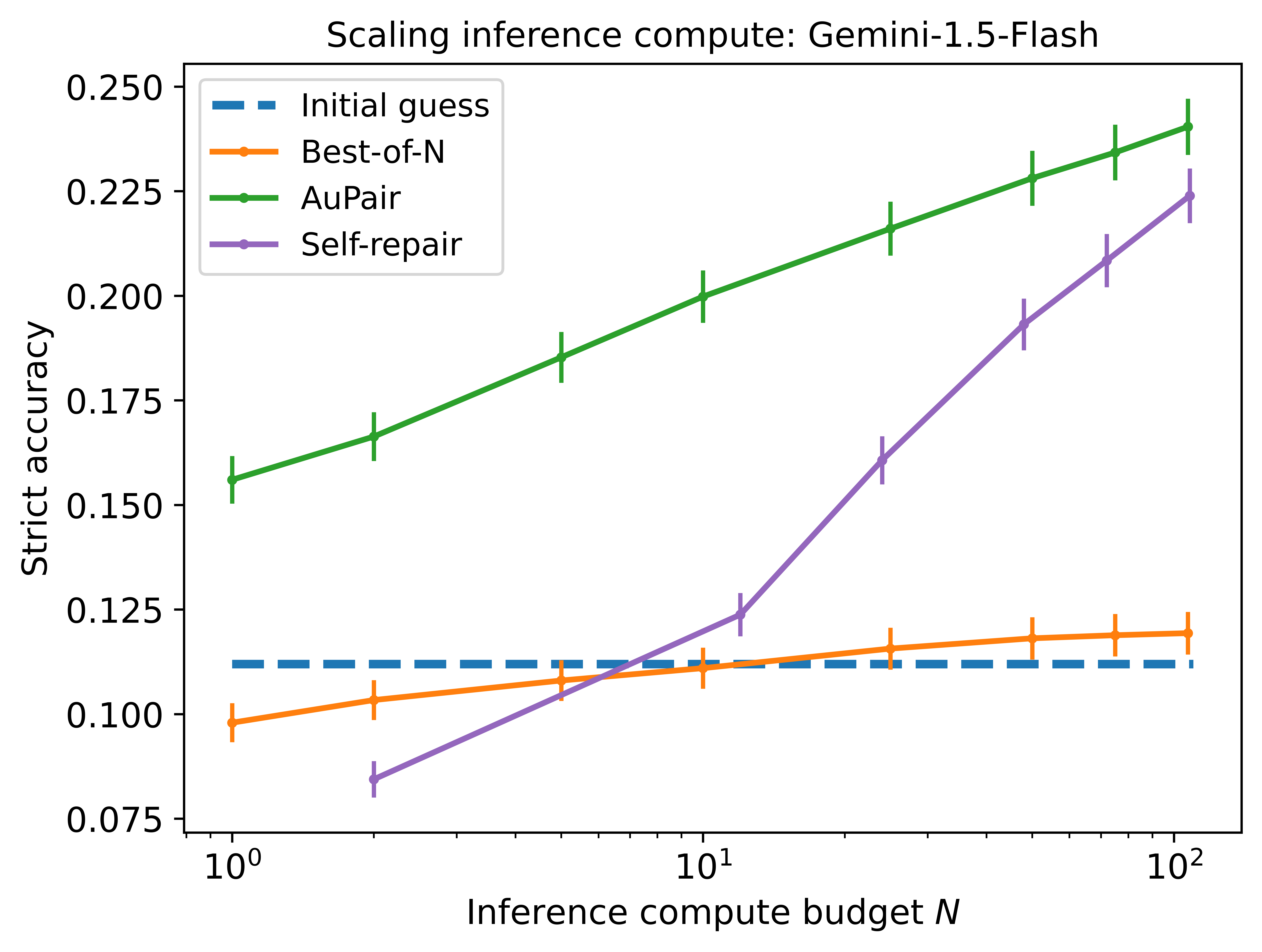}
    \caption{\textbf{Strict accuracy when scaling inference-time compute}: with $N = 144$ for Gemini-1.5-Pro and $N = 110$ for Gemini-1.5-Flash}
    \label{fig:inference_compute_scaling_strict_accuracy}
\end{figure}

In addition to the scaling experiment we performed using Gemini-1.5-Pro (results in Fig.~\ref{fig:pairs_vs_aupairs}(b)), we also perform the same scaling experiment using Gemini-1.5-Flash and show the results in Fig.~\ref{fig:inference_compute_scaling_flash}. Moreover, we report the results of the same scaling experiment on the strict accuracy metric in Fig.~\ref{fig:inference_compute_scaling_strict_accuracy}. The trend is similar to what we observed before: best-of-$N$ plateaus after a certain number of LLM calls, while our approach scales as the compute budget increases, delivering an improvement in performance for each newly included \aupair/. The self-repair baseline performs better with the Gemini-1.5-Flash model than with the Pro model; our hypothesis is that since the initial guesses for the Pro model were worse because of formatting issues, self-repair did not yield significant improvements. However, when the initial guesses are better, the self-repair baseline shows a stronger scaling result. Our algorithm yields 110 \aupairs/ and achieves a final test pass rate of 37.83\% and strict accuracy 24.14\%. Best-of-$N$, on the other hand, given the same budget of 110 LLM calls, has a test pass rate of 21.8\% and strict accuracy 11.93\%. Self-repair with the same compute budget has a final test pass rate of 34.1\% and strict accuracy 22.39\%. Since our \aupairs/ are selected submodularly, the initial pairs yield high returns in performance and these returns start diminishing slowly, but notably, performance does not plateau yet. Thus, it is abundantly clear that using \aupairs/ has a distinct advantage over currently used approaches like best-of-$N$ and self-repair in improving performance at inference time as compute budget increases.

\begin{figure}
    \centering
    \includegraphics[width=0.45\linewidth]{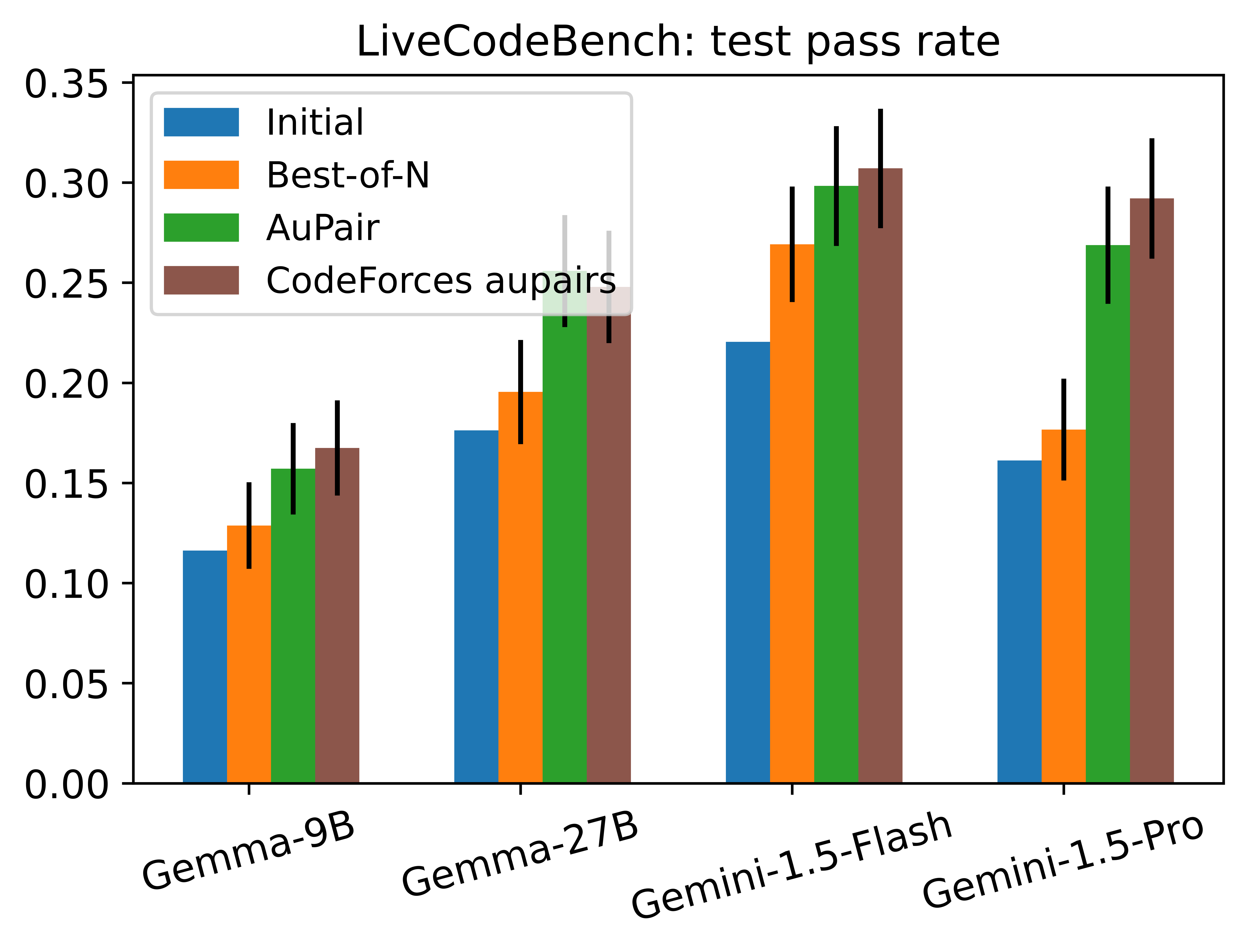}
    \includegraphics[width=0.45\linewidth]{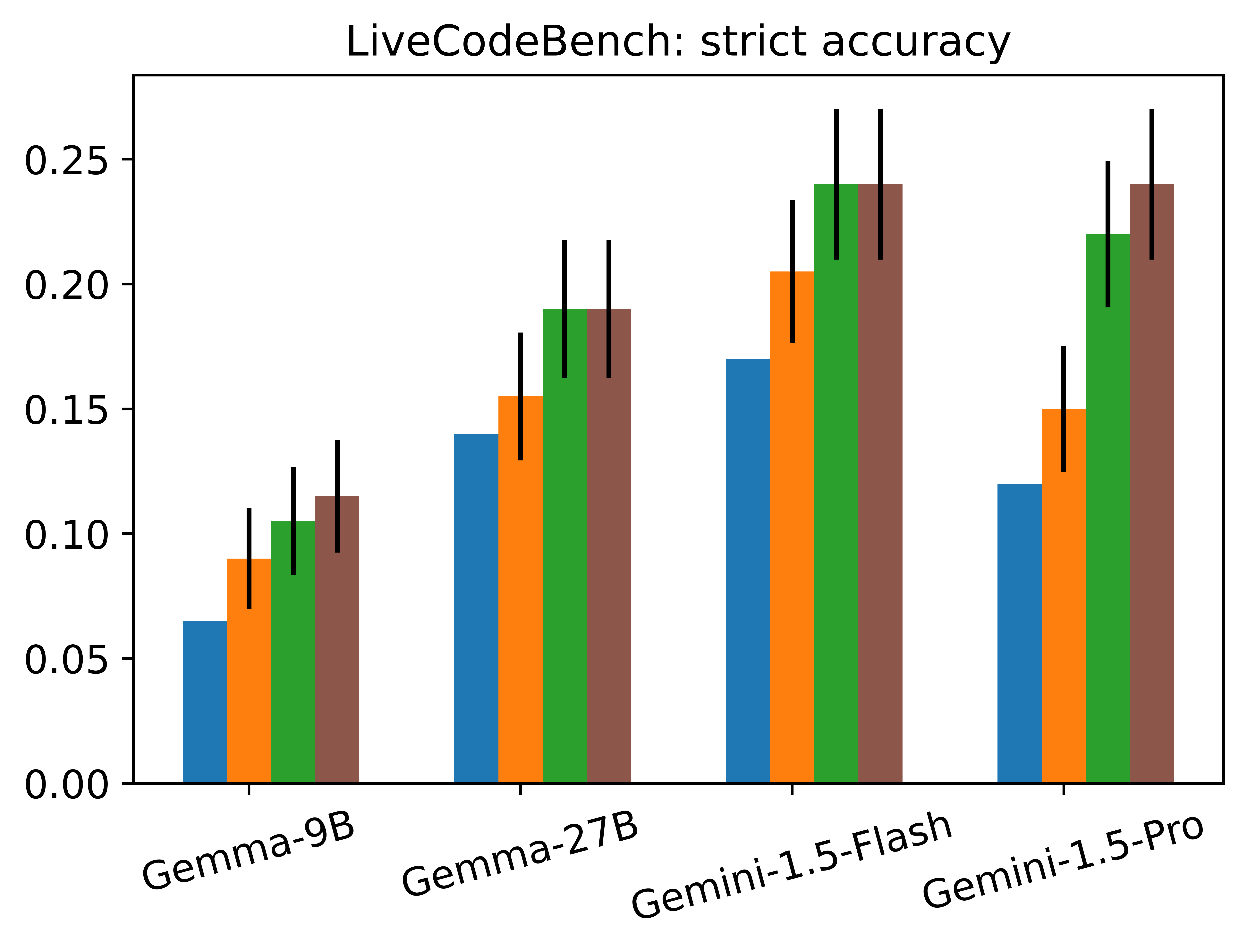}
    \caption{\textbf{LiveCodeBench results}: using \aupairs/ from the CodeForces dataset matches or outperforms in-distribution \aupairs/ from LiveCodeBench (left: test pass rate, right: strict accuracy).}
    \label{fig:lcb}
\end{figure}

\subsection{Code Repair with LiveCodeBench}\label{sec:LCB}

Generalisation of \aupair/ prompting is important to improve code repair of smaller datasets. We posit that the \aupairs/ contain diverse code changes that transfer meaningfully across datasets, which may be important to those with scarce data, since out-of-distribution generalisation becomes especially relevant when we have small datasets, on which it can be quite difficult to obtain many different \aupairs/.

We now show the results obtained for a smaller dataset (400 problems) LiveCodeBench (LCB)~\citep{jain2024livecodebench}. We generate the same train/val/test split (37.5/12.5/50$\%$) over 400 problems and apply our \aupair/ approach to obtain in distribution \aupairs/ for LCB. 

Fig.~\ref{fig:lcb} shows that even with smaller number of selected \aupairs/ we still obtain a gain over best-of-$N$ prompting. We obtained 5 \aupairs/ with the submodular extraction in Algorithm~\ref{alg:submodular} for all the models except Gemma-9B which obtained only 3 \aupairs/. Given the difference in dataset size these values are larger in proportion to the ones obtained from a larger dataset CodeForces (8.8k problems, 144 extracted \aupairs/).

Another interesting result in Fig.~\ref{fig:lcb} is that both metrics, the test pass rate and strict accuracy, are comparable when using in-distribution \aupairs/ from LiveCodeBench and out-of-distribution \aupairs/ from CodeForces. This reinforces the insight mentioned earlier that extracting \aupairs/ on one dataset could lead to significant improvements over baselines even on other datasets.

\subsection{Lineage}

\begin{figure}
    \centering
    \includegraphics[width=\linewidth]{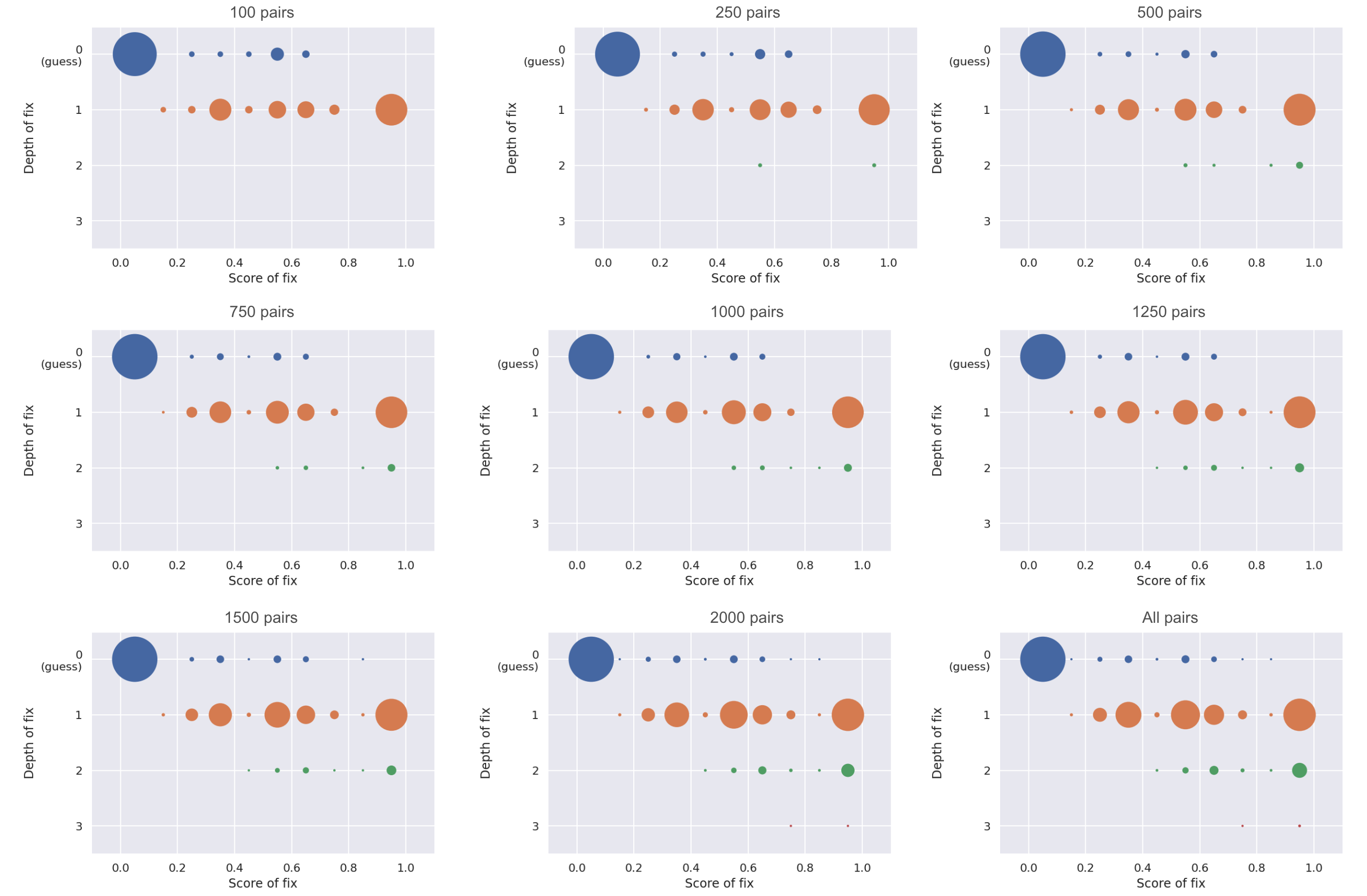}
    \caption{Visualising the lineage of the set of all pairs as the first phase of the algorithm, pair generation, progresses.}
    \label{fig:ancestry_evolution}
\end{figure}

Here we look at the lineage of each pair generated during phase 1 of our algorithm, pair generation. The key idea here is to see if the set of all pairs collected during the pair generation phase are \emph{deeper} i.e., they generate iteratively better solutions for a smaller set of problems, or \emph{broader} i.e., they generate solutions for a larger set of problems but those solutions may not necessarily be perfect. The last plot in Fig.~\ref{fig:ancestry_evolution} (pairs generated on the CodeForces dataset using Gemini-Pro-1.5) indicates that the pairs collected have shallow lineage: a large proportion of guesses that had a score of 0 had corresponding fixes with perfect score at depth 1. We also see that the number of fixes decreases as depth increases (as seen from the size of the circles), indicating that several problems could not be improved beyond a certain point, or that they were not resampled during the pair generation phase. In both these cases, one solution is to allow more LLM calls during phase 1 to allow each problem to be sampled for repair more times. The takeaway here is that more sophisticated fixes for difficult problems can be discovered as we increase the budget of LLM calls during the pair generation phase. The entire evolution of this lineage at different points during pair generation is illustrated in Fig.~\ref{fig:ancestry_evolution}.

\subsection{Code Diversity}\label{sec:appendix_diversity}

We compute the code diversity score in Fig.~\ref{fig:cross_model_transfer}(b) based on the number of different abstract syntax subtrees that each code instance produces. Algorithm~\ref{alg:diversity} describes how this diversity score is computed. As a first step, for each guess in the test dataset and its corresponding fix generated by the LLM, we compute the respective abstract syntax subtrees. Next, we compute their corresponding set difference to get the unique subtrees for each code diff. This is done $N$ times for a compute budget of $N$ and the number of code diff subtrees across pairs and problems is averaged and normalised in the following manner to yield the diversity score $\delta$:

\begin{align}
    \delta = \frac{1}{N |\mathcal{D}_{\text{test}}| |S_\text{max}|} \sum \limits_{i=1}^{|\mathcal{D}_{\text{test}}|} |S^{\text{diff}}_i|
\end{align}

where $S^{\text{diff}}_i$ is the set of all code diff subtrees generated with compute budget $N$ for problem $i$, and the normalising factor $S_\text{max}$ corresponds to the highest number of subtrees that are present in any such set of code diff subtrees.

\begin{tabular}{c}
\begin{minipage}[c]{\linewidth}     
\begin{algorithm}[H]
\begin{algorithmic}[1]
\Require
$\left\{\begin{array}{ll}
N & \text{inference compute budget}\\
\mathcal{D_\text{test}} & \text{test dataset}\\
\hat{\mathcal{Y}} & \text{fixes for test problems}\\
f_\text{AST} & \text{abstract syntax subtree computation function} \\
\end{array}
\right.$
\State init set of code diff subtrees: $S^\text{diff} \gets []$
\For{problem $\bm{x} \in \mathcal{D}_\text{test}$ and its corresponding fixes $\bm{\hat{y}} \in \hat{\mathcal{Y}}$}
\State compute abstract syntax subtrees for guess: $S^{\text{guess}} \gets f_{\text{AST}}(\bm{x}^{\text{guess}})$
\State init code diff subtrees for this problem: $s \gets \emptyset$
\For{$j \in \{1, \ldots, N\}$}
\State compute abstract syntax subtrees for fix: $S^{\text{fix}}_j \gets f_{\text{AST}}(\hat{\bm{y}}_j)$
\State update code diff subtrees: $s \gets s \cup \{S^\text{fix}_j \setminus S^\text{guess}\}$
\EndFor
\State{append to code diff subtrees: $S^\text{diff} \gets S^\text{diff} + s$}
\EndFor
\State{compute normalising factor: $S_\text{max} \gets \text{argmax}_i S^\text{diff}_i$}
\State{compute diversity score: $\delta \gets \frac{1}{N |\mathcal{D}_\text{test}| |S^\text{max}|} \sum \limits_{i=1}^{|\mathcal{D}_{\text{test}}|} |S^{\text{diff}}_i|$}
\Statex \Return{$\delta$}
\end{algorithmic}
\caption{Diversity score computation}
\label{alg:diversity}
\end{algorithm}
\end{minipage}
\end{tabular}

\subsection{Prompting}\label{sec:prompting}

There are 2 types of prompts that we use: 1) guess generation prompt, and 2) repair prompt. The guess generation prompt is used during dataset creation, for obtaining the initial guesses for all problems in the dataset. The repair prompt is used throughout the rest of the paper: in the Pair Generation (Phase 1,~\S\ref{sec:phase1} with $k=32$ random examples) and in the \aupair/ Extraction (Phase 2,~\S\ref{sec:phase2}) and during inference, with $k=1$. The function signature indicates that the function expects a string as an input. The instruction specifies that the final answer is meant to be printed \emph{inside} the function, and that the main function is not meant to be written.

The structure of our repair prompt is as follows: there is an instruction at the top, followed by the few-shot examples in the format: question, guess, fix. We also add the score achieved by the guess and the fix for the in-context example pairs. Following this, we add the text and initial guess for the problem and the LLM then has to generate a better fix.  Note that we do not provide any extra execution feedback in the form of execution traces; this could potentially be explored by future work.

\begin{tcolorbox}[colback=blue!5!white,colframe=black,title= Guess Generation Prompt,fonttitle=\bfseries]

$\texttt{\footnotesize <problem text>}$

Complete the function definition below. Print the final answer in the function. Do not write main. Do not write anything outside the $\texttt{\footnotesize solve()}$ function. 

\begin{lstlisting}[language=Python,basicstyle=\footnotesize\ttfamily]
def solve(s: str):
  ...
\end{lstlisting}

\end{tcolorbox}

\begin{tcolorbox}[colback=blue!5!white,colframe=black,title= Repair Prompt,fonttitle=\bfseries]\label{fig:repair_prompt}
You are an experienced software developer.

Look at the question (Q) and solutions below (A).

The main objective is to improve the $\texttt{\footnotesize solve()}$ function to answer the question.\\

Example $\texttt{1}$:\\

(Q): ...

Bad solution code $\texttt{\footnotesize A(bad)}$:

\begin{lstlisting}[language=Python,basicstyle=\footnotesize\ttfamily]
def solve(s: str):
  ...
\end{lstlisting}

The score of this code is $\texttt{\footnotesize score(A(bad)) =  <example\_guess\_score>}$.\\

Good solution code $\texttt{\footnotesize A(good)}$:

The score of this code is $\texttt{\footnotesize score(A(good)) =  <example\_fix\_score>}$.

\begin{lstlisting}[language=Python,basicstyle=\footnotesize\ttfamily]
def solve(s: str):
  ...
\end{lstlisting}

$\vdots$\\

=======================================\\

The main objective is to improve the $\texttt{\footnotesize solve()}$ function to answer the question.

(Q): ...

Bad solution code $\texttt{\footnotesize A(bad)}$:

\begin{lstlisting}[language=Python,basicstyle=\footnotesize\ttfamily]
def solve(s: str):
  ...
\end{lstlisting}

The score of this solution is $\texttt{\footnotesize score(A(bad)) =  <guess\_score>}$\\

Good solution code $\texttt{\footnotesize A(good)}$:

The score of this solution is $\texttt{\footnotesize score(A(good)) = 100}$
\\

\end{tcolorbox}

\begin{wrapfigure}[11]{R}{0.5\linewidth}
    \centering
    \includegraphics[width=\linewidth]{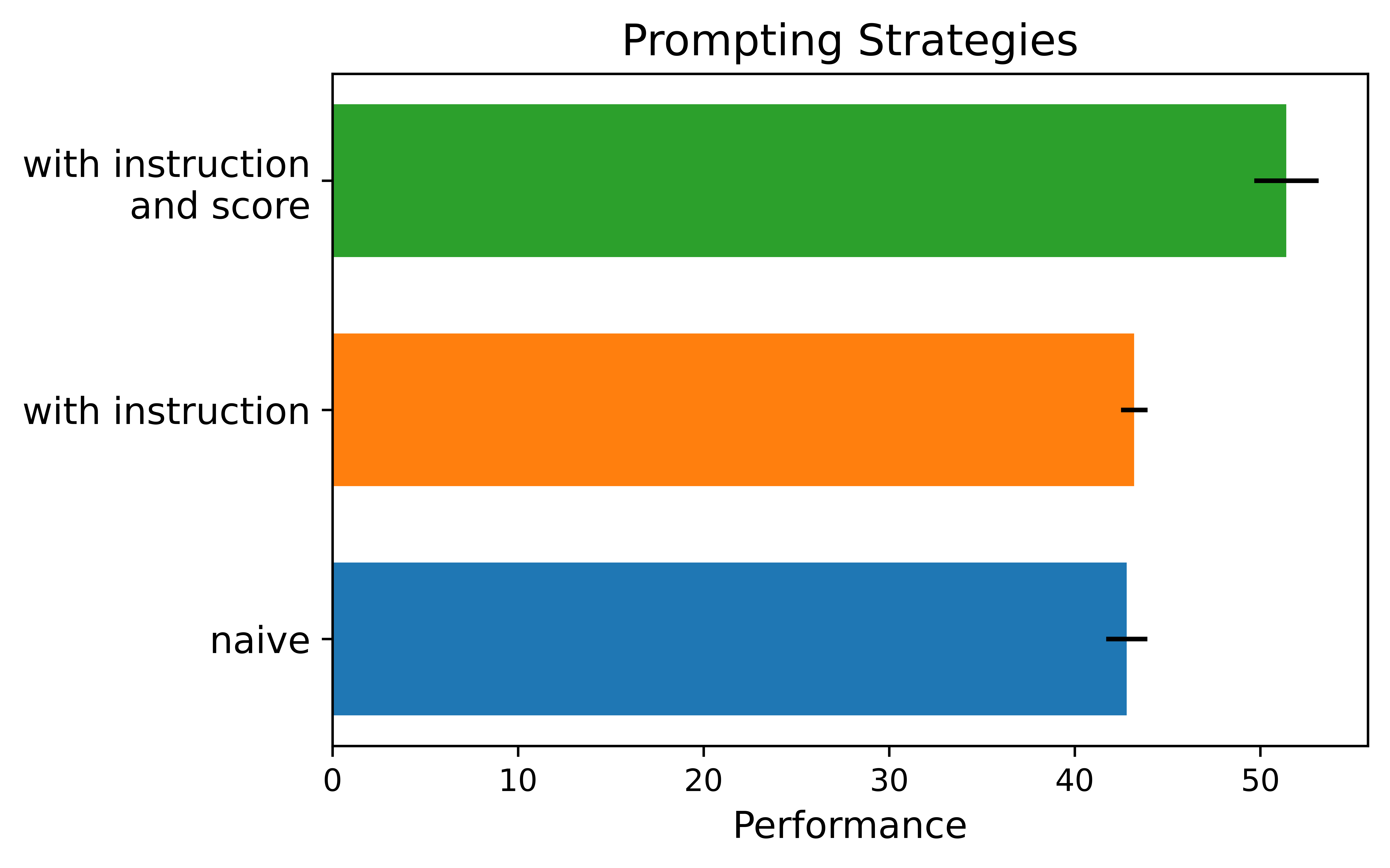}
    \label{fig:prompting_strategies}
\end{wrapfigure}

Our aim is clear: the pairs indicate a certain type of change and we provide these pairs in context to aid the LLM in generating an improved solution for the given problem. Some different prompting strategies that we tried out were the following:

\textit{Na\"ive prompting}: only include the problem, guess and fix for the pairs, followed by the problem and guess for the test problem.

\textit{Prompting with instruction only}: include the header instruction followed by the components of the na\"ive prompting strategy.

\textit{Prompting with instruction and score}: include the elements of 2 above, but in addition, also include the score that each guess and fix received on the corresponding problem's test cases. This is the prompt that we finally use and the one that gives us better results when compared using the same set of pairs with the previous 2 strategies. An important thing to note here is that we prompt the model with a desired fix score of 100 for the test problem.

We test the three strategies described above on a subset of the CodeForces dataset and report their performance in terms of number of problems solved, in the figure on the right. The results clearly indicate that the final prompting strategy that includes the instruction and score is the best strategy and so we choose it to compose the repair prompt.

\subsection{Code Execution}\label{sec:code_execution}

When the LLM generates a fix for any problem, we call the $\texttt{\footnotesize solve()}$ function for each test case associated with that problem. We then compare the output with the ground truth and give a partial score corresponding to the proportion of test cases passed by this fix.

An important point to note is that the $\texttt{\footnotesize solve()}$ function has to take as input a string, which is then parsed into the correct variables. This formatting requirement is a key reason for the poor initial performance of Gemini-1.5-Pro in Fig.~\ref{fig:in_dist_performance}. Since the instruction for generating the initial guess is not correctly followed by the model, a lot of guesses end up invariably having incorrect parsing of the input, leading to low scores. A lot of \aupairs/ extracted using these models, as a result, contain this formatting fix, as we will see in \S\ref{sec:actual_aupairs}.

\subsection{Types of Fixes in \aupairs/}\label{sec:actual_aupairs}

We now show some examples of \aupairs/ and highlight the differences between the guess and fix for each pair. These are a mix of CodeForces pairs collected using different models. The scores achieved by the guess and fix on the corresponding problem's test cases are specified at the top right corner for each example in Fig.~\ref{vis:aupairs}. We also provide a short description for each type of fix in the caption for ease of understanding. The types of pairs discovered using our algorithm cover a large area of potential fixes that can be made to an initial buggy piece of code: from smaller ones like parsing, fixing logical bugs pertaining to indexing errors, variable initialisations, etc., to larger changes like rewriting parts of the code, or even suggesting alternate routes to solve the same problem.

\begin{figure}[!b]
    \centering

    \begin{subfigure}{\linewidth}
    \includegraphics[trim={1cm 5cm 3cm 3cm}, clip, width=\linewidth]{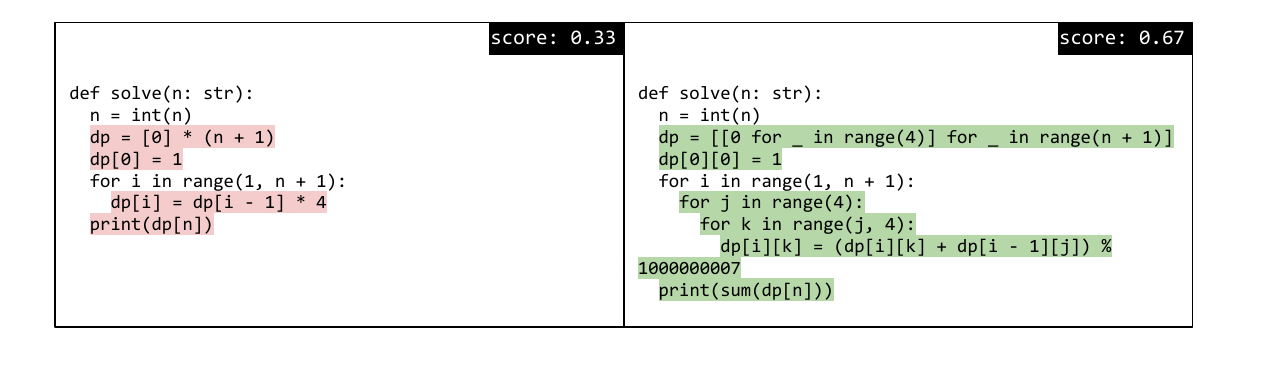}
    \subcaption{Fix: converts a 1-D dynamic programming solution to 2-D.}
    \end{subfigure}
    \medskip
    
    \begin{subfigure}{\linewidth}
    \includegraphics[trim={1cm 1.5cm 3cm 3cm}, clip, width=\linewidth]{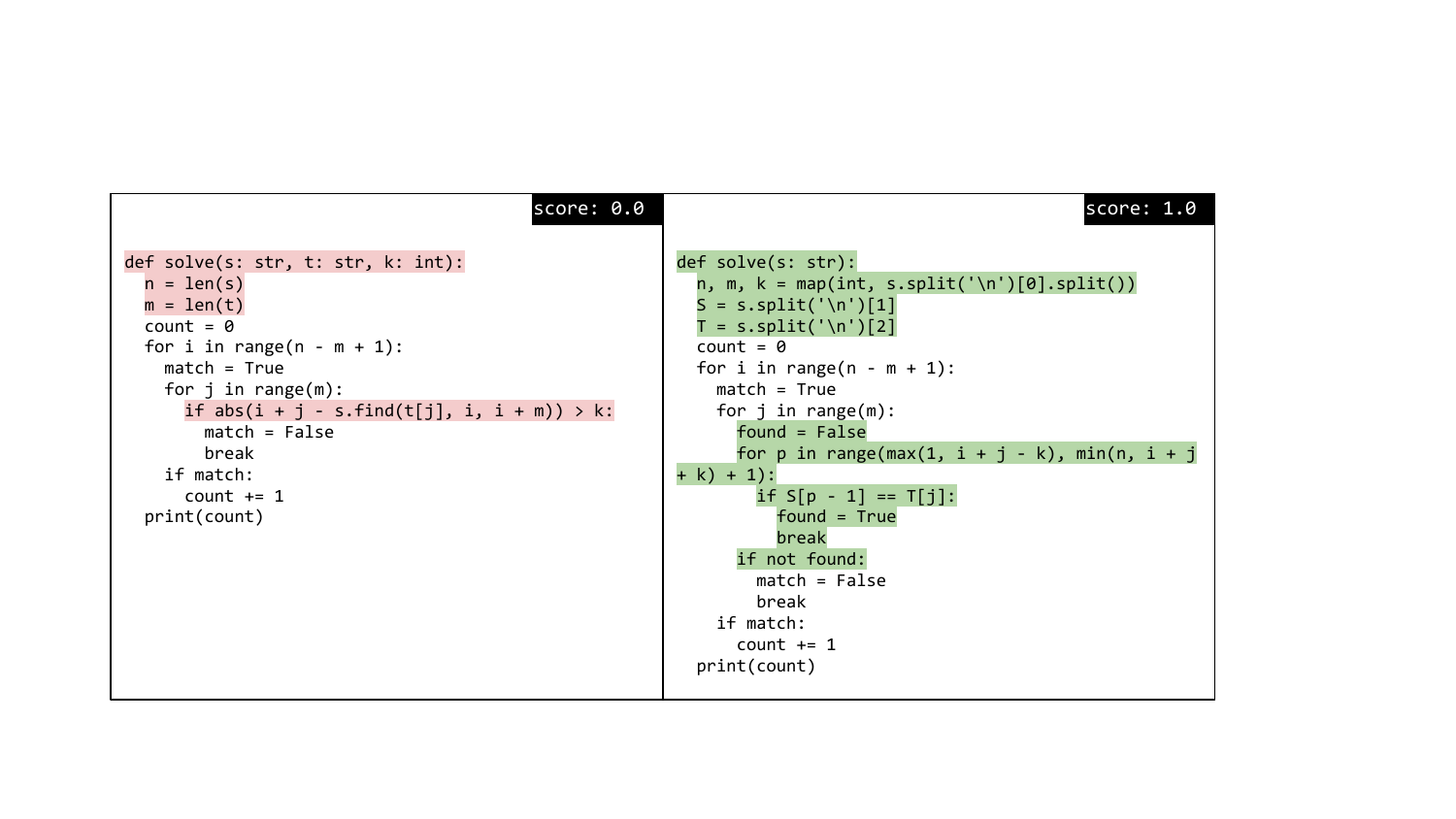}
    \subcaption{Fix: composition of 1) input parsing correction, and 2) logical bug fix.}
    \end{subfigure}
    \medskip
    
    \begin{subfigure}{\linewidth}
    \includegraphics[trim={1cm 4cm 3cm 3cm}, clip, width=\linewidth]{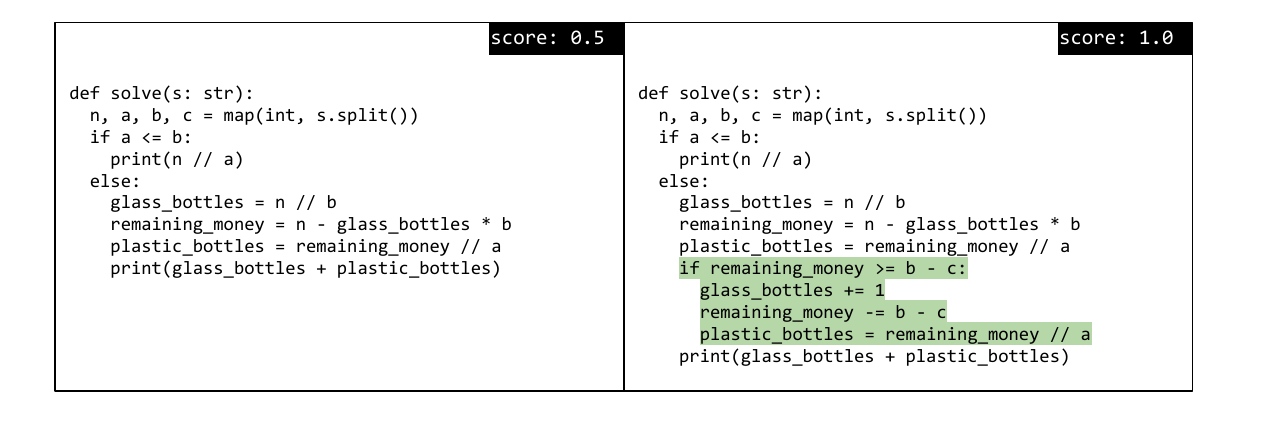}
    \subcaption{Fix: add an extra condition for edge cases.}
    \end{subfigure}
    \medskip
\end{figure}

\begin{figure}[ht]\ContinuedFloat
    \centering
    \begin{subfigure}{\linewidth}
    \includegraphics[trim={1cm 4cm 3cm 3cm}, clip, width=\linewidth]{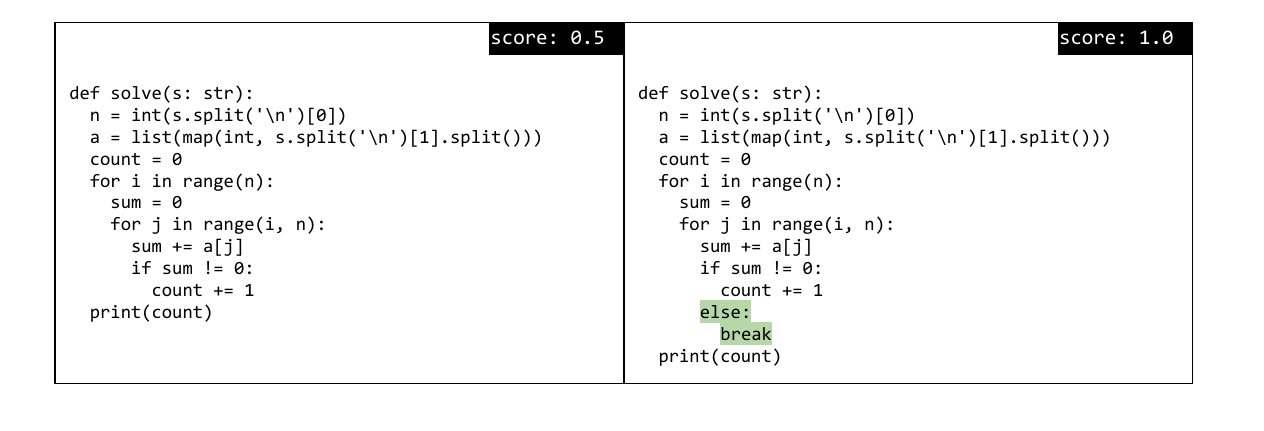}
    \subcaption{Fix: add loop exit condition.}
    \end{subfigure}
    \medskip
    
    \begin{subfigure}{\linewidth}
    \includegraphics[trim={1cm 4cm 3cm 3cm}, clip, width=\linewidth]{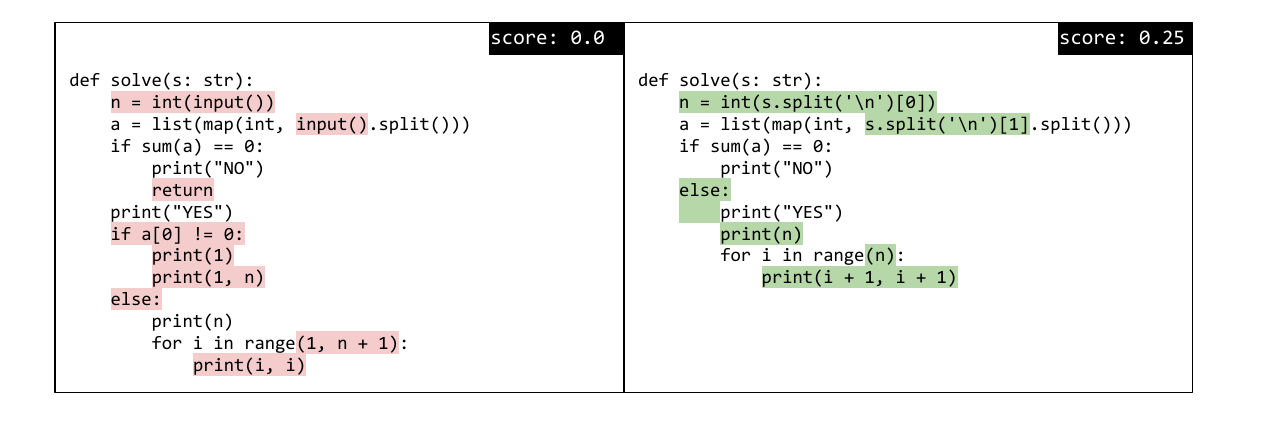}
    \subcaption{Fix: composition of 1) input parsing correction, and 2) logical bug fix.}
    \end{subfigure}
    \medskip
    
    \begin{subfigure}{\linewidth}
    \includegraphics[trim={1cm 2cm 3cm 3cm}, clip, width=\linewidth]{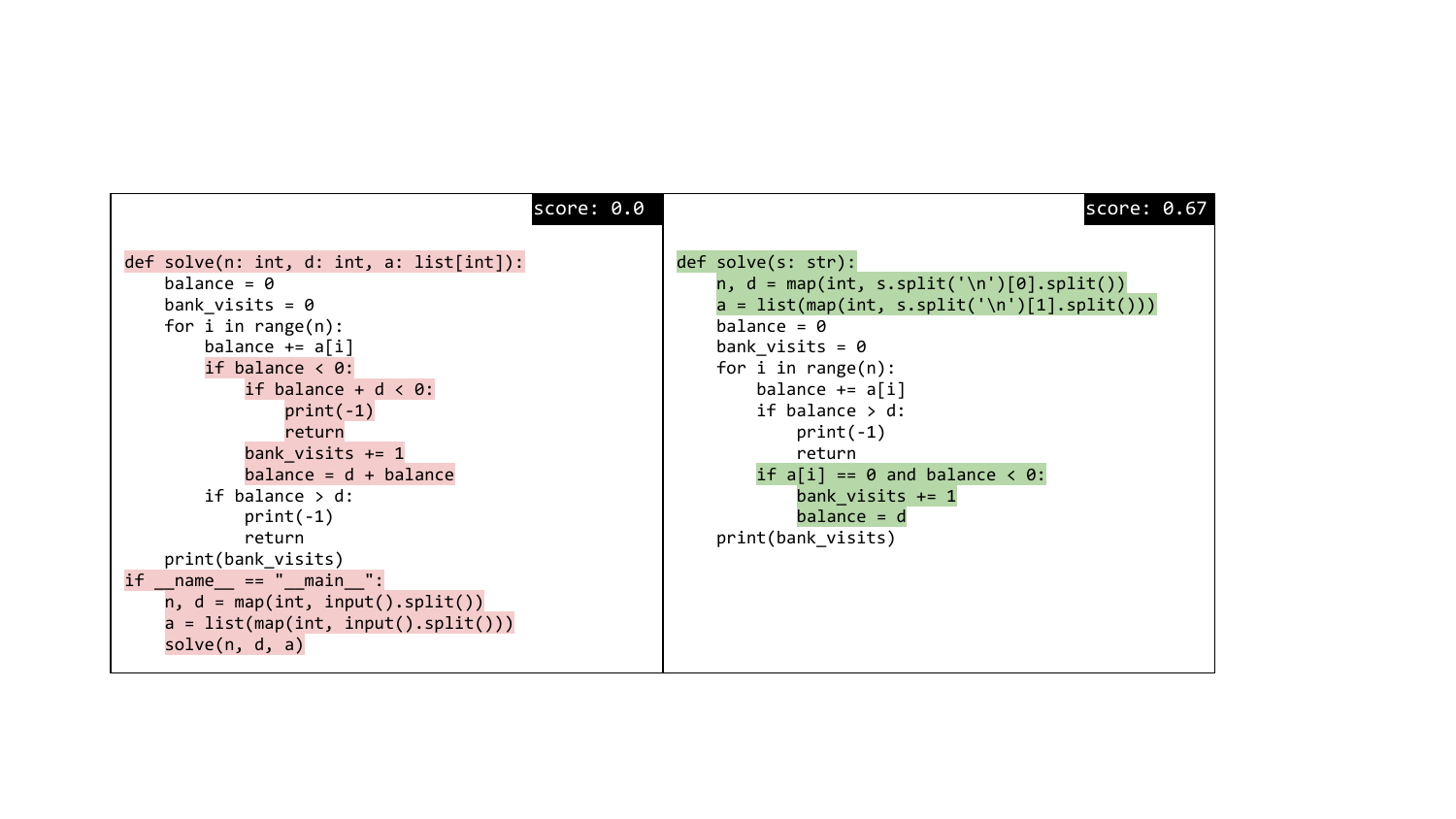}
    \subcaption{Fix: composition of 1) function signature correction, 2) input parsing correction, and 3) logical bug fix.}
    \end{subfigure}
    \medskip
    
\end{figure}

\begin{figure}[ht]\ContinuedFloat
    \centering
    \begin{subfigure}{\linewidth}
    \includegraphics[trim={1cm 6cm 3cm 3cm}, clip, width=\linewidth]{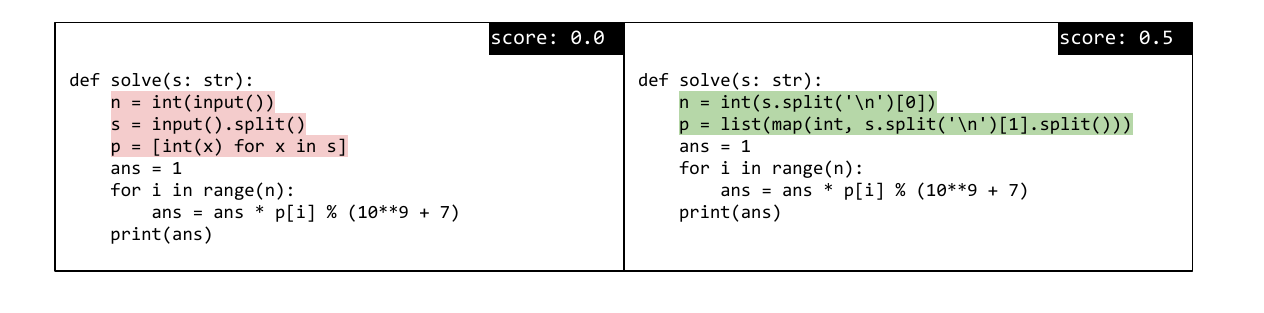}
    \subcaption{Fix: input parsing correction.}
    \end{subfigure}
    \medskip

    \begin{subfigure}{\linewidth}
    \includegraphics[trim={1cm 4cm 3cm 3cm}, clip, width=\linewidth]{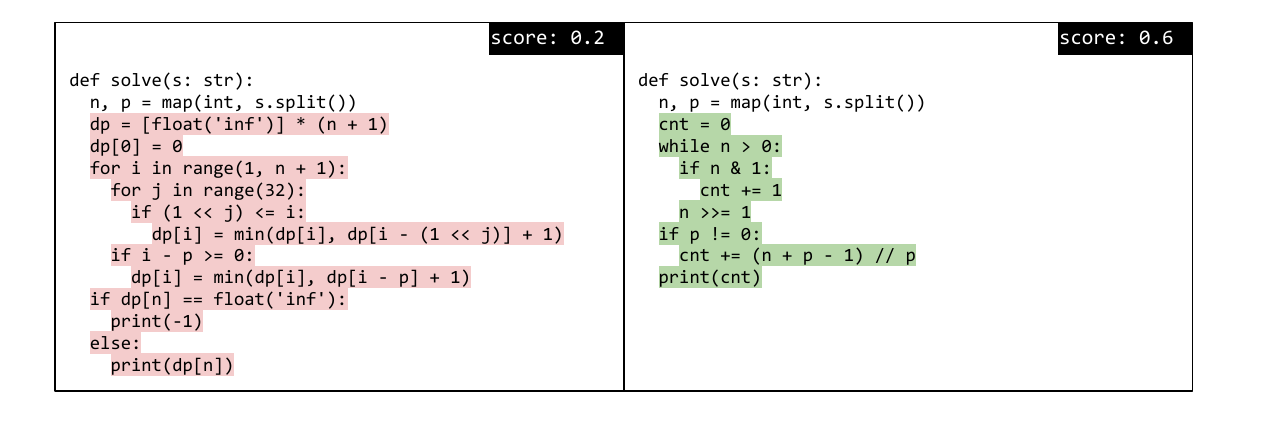}
    \subcaption{Fix: solve problem using bit manipulation instead of dynamic programming.}
    \end{subfigure}
    \medskip
    
    \begin{subfigure}{\linewidth}
    \includegraphics[trim={1cm 1cm 3cm 1cm}, clip, width=\linewidth]{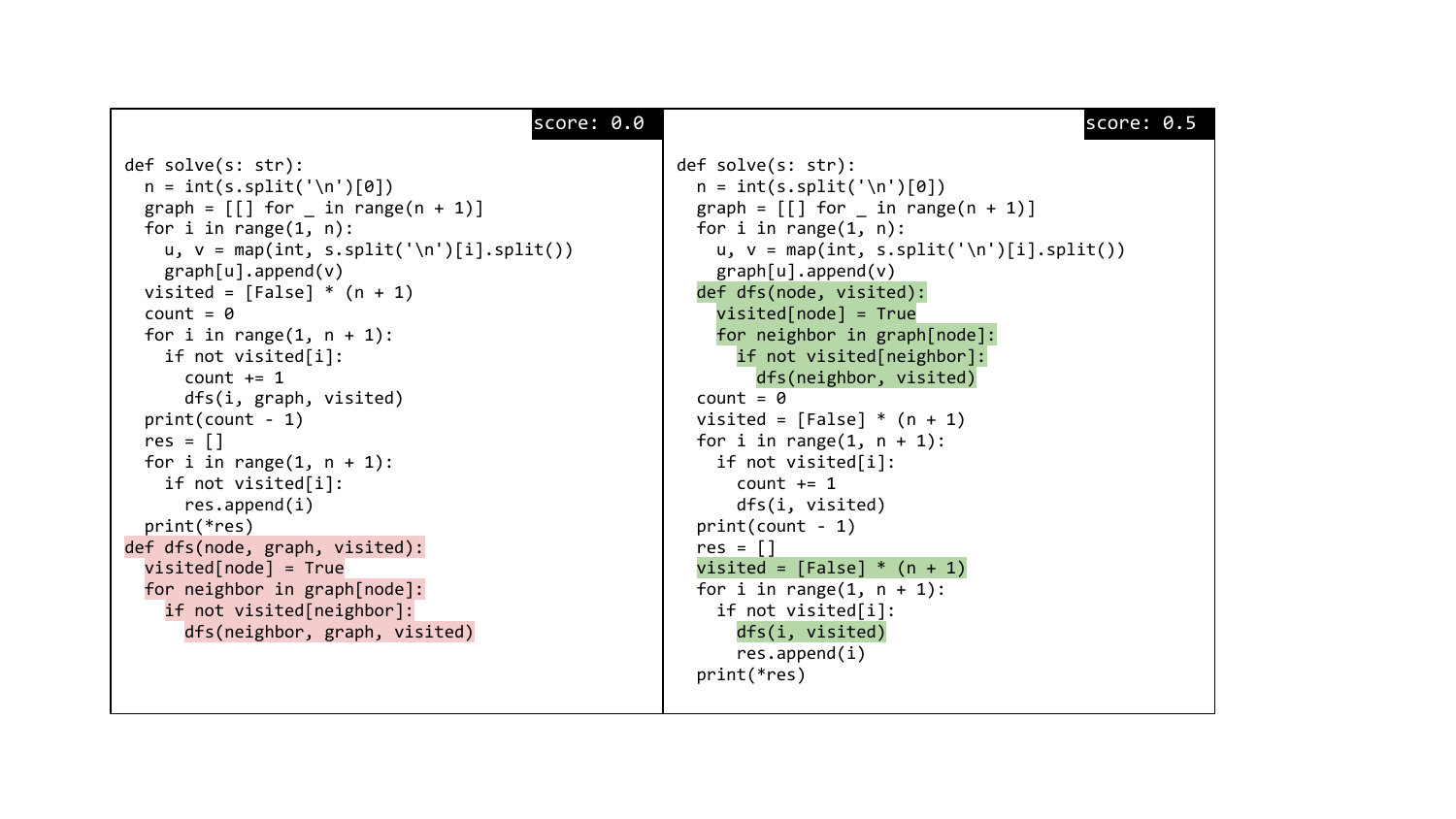}
    \subcaption{Fix: partial correction to depth-first search graph algorithm.}
    \end{subfigure}
    \medskip
    
\end{figure}

\begin{figure}[ht]\ContinuedFloat
    \centering
    \begin{subfigure}{\linewidth}
    \includegraphics[trim={1cm 1cm 3cm 3cm}, clip, width=\linewidth]{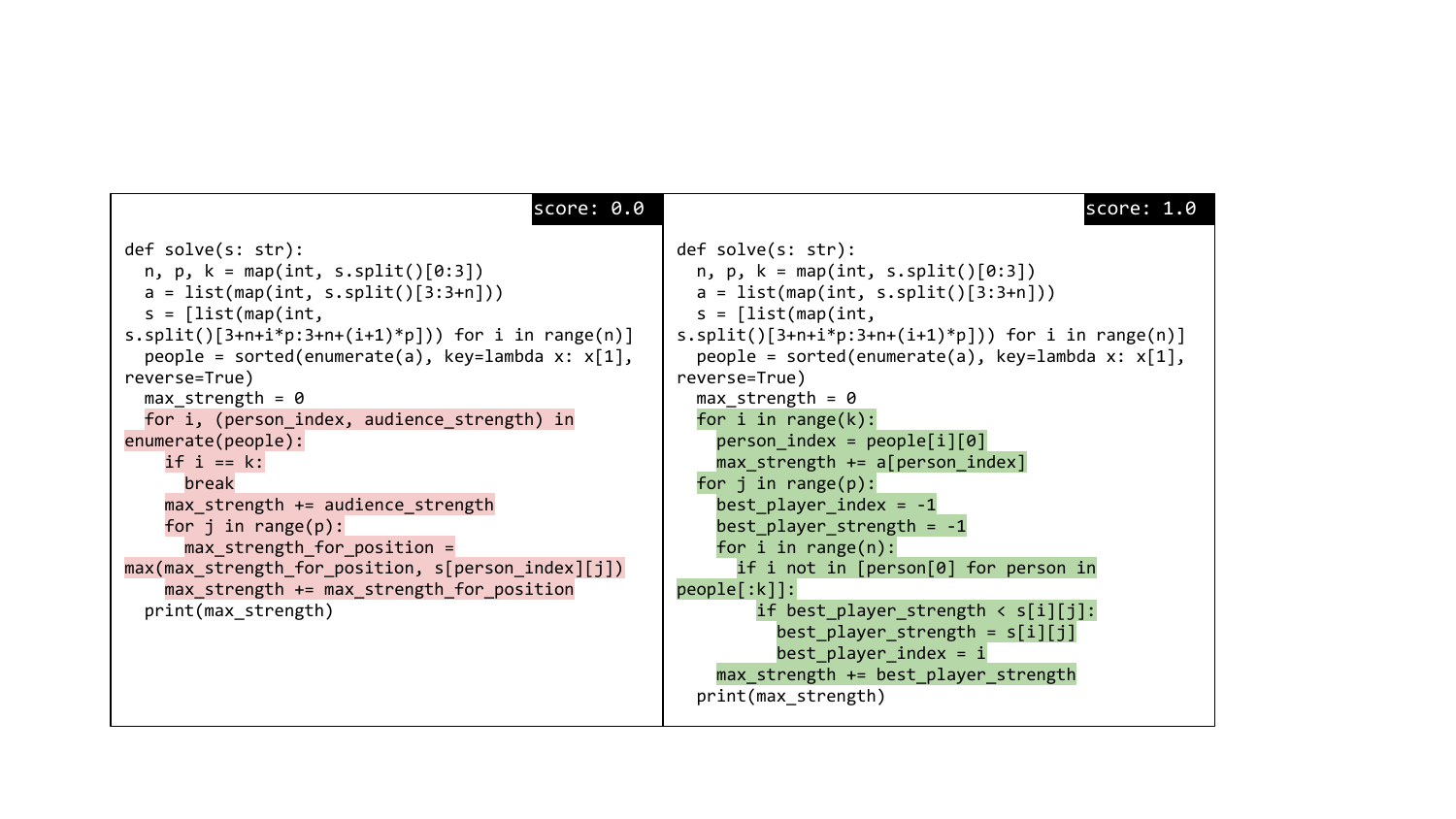}
    \subcaption{Fix: rewrite partial solution to pass all test cases.}
    \end{subfigure}
    \medskip
    \caption{Examples of \aupairs/ produced by our algorithm (multiple models represented above)}
    \label{vis:aupairs}
\end{figure}

\end{document}